\documentclass[a4paper,fleqn,usenatbib]{mnras}

\usepackage[T1]{fontenc}
\usepackage{ae,aecompl}

\usepackage{graphicx}	
\usepackage{amsmath}	
\usepackage{amssymb}	

\title{Formation of satellites in circumplanetary discs generated
by disc instability}

\author[C. Inderbitzi et al.]{
	C. Inderbitzi$^{1}$\thanks{E-mail: cassandra.inderbitzi@uzh.ch},
	J. Szul\'{a}gyi$^{1}$,
	M. Cilibrasi$^{1}$,
	L. Mayer$^{1}$
	\\
	$^{1}$Center for Theoretical Astrophysics and Cosmology, Institute for Computational Science, University of Z\"{u}rich,\\
	Winterthurerstrasse 190, CH-8057 Z\"{u}rich, Switzerland\\
}

\date{Accepted XXX. Received YYY; in original form ZZZ}

\pubyear{2019}

\begin{document}
	\label{firstpage}
	\pagerange{\pageref{firstpage}--\pageref{lastpage}}
	\maketitle
	
	\begin{abstract}
	We investigated the formation and evolution of satellite systems in a cold, extended circumplanetary disc around a 10 $M_{\rm{Jupiter}}$ gas giant which was formed by gravitational instability at 50\,AU from its star. The disc parameters were from a 3D global SPH simulation. We used a population synthesis approach, where we placed satellite embryos in this disc, and let them accrete mass, migrate, collide until the gaseous disc is dissipated. In each run we changed the initial dust-to-gas ratio, dispersion- and refilling time-scales within reasonable limits, as well as the number of embryos and their starting locations. We found that most satellites have mass similar to the Galilean ones, but very few can reach a maximum of 3 $M_{\rm{Earth}}$ due to the massive circumplanetary disc. Large moons are often form as far as 0.5 $R_{\rm{disc}}$. The migration rate of satellites are fast, hence during the disc lifetime, an average of 10 $M_{\rm{Earth}}$ worth of moons will be engulfed by the planet, increasing greatly its metallicity. We also investigated the effect of the planet's semi-major axis on the resulting satellite systems by re-scaling our model. This test revealed that for the discs closer to the star, the formed moons are lighter, and a larger amount of satellites are lost into the planet due to the even faster migration. Finally, we checked the probability of detecting satellites like our population, which resulted in a low number of $\leq$ 3\% even with upcoming powerful telescopes like E-ELT.
	\end{abstract}
	
	
	
	
	\section{Introduction}
	
	There are two major scenarios of gas giant planet formation, core-accretion (CA; \citealt{Pollack96}) and gravitational instability  (GI; \citealt{Boss97},\citealt{Mayer02}). In both formation mechanisms a circumplanetary disc (CPD; e.g. \citealt{Lubow99}, \citealt{Kley99}, \citealt{Shabram13}) is created at the last stage of the formation process. Furthermore, in a similar way as planets form within the circumstellar disc, satellites can form in these CPDs \citep{Lunine82}. So far, a comparative study of the properties of the satellite systems formed in CPDs arising in CA or GI has never been attempted.

    In our Solar System, every gas- and ice giant has a satellite system. However, we only have one possible exomoon candidate around an exoplanet so far \citep{Teachey18}. Of course, detecting exomoons is particularly challenging, not only because of their tiny mass and size. It is also because a priori, CPD forming massive planets are created beyond the snowline and these are the ones which can have satellite systems. Most exoplanet/exomoon detection methods are sensitive to planets much closer in. While dynamical interactions could bring giant planets closer to the star, during these some/all moons could detach from the planet or be torn apart in tidal interactions. Therefore, detecting exomoons is particularly challenging as the bulk of the population is orbiting in the outer planetary systems.

    To understand the possible differences between satellite systems that form in GI or CA CPDs, the characteristics of these discs have to be taken into account. In the GI case, previous works have found massive CPDs: \citet{Galvagni12} with disc masses of 0.5 $M_{\rm{planet}}$, while \citet{Shabram13} reported 0.25 $M_{\rm{planet}}$. However, given that the CPD is fed by the circumstellar disc through the meridional circulation \citep{Szulagyi14,Fung16}, it is natural that the CPD mass will scale with the CSD mass. On top of that, for GI to operate one needs a very massive CSD, which further contributes to the massive CPDs. It was shown by \citet{Szulagyi17} that the CPD mass indeed linearly scales with the CSD mass, and it also scales with the planet mass. Regarding the temperatures of CPDs, the GI scenario predicts very low values: $\sim$100 K or below  \citep{Galvagni12,Shabram13}.
    CPDs that formed during the CA scenario were studied more frequently. The subdisc masses were spanning a large range from $10^{-4}$ to $10^{-2}$ planet masses \citep{Dangelo03,Gressel13,Szulagyi14,Szulagyi16b,Szulagyi17}, since again, the CPD mass will depend on the planet mass and the CSD mass. The temperatures, in works employing radiative simulations, were significantly higher than those of CPDs in GI simulations. Furthermore, within the CPD there was a large temperature gradient, ranging from several thousand Kelvin near the planet, to a few hundred Kelvin at the edge of the CPD \citep{AyliffeBate09,Szulagyi16a,Szulagyi17,SZM17}. 
    To understand the real differences between GI and CA CPDs, \citet{Szulagyi16b} compared the masses and temperatures within two simulations for the two cases. The initial parameters of these simulations were the same, the planets were of 10 $\rm{M_{\rm{Jupiter}}}$ at 50 AU from their stars. While the CA CPD was only 8 times smaller in mass than the GI CPD, their temperatures were an order of magnitude different, with the CA CPD being significantly hotter. This naturally rises the question of how different would be the satellite system formed from such markedly different conditions.

    So far, satellite formation was mainly studied by population synthesis. \citet{Sasaki10} has investigated the outcome of Saturnian and Jovian moon systems based on the differences in disc mass, cavity between the disc and planet, as well as the termination time of gas infall. \citet{Miguel16} used a minimum-mass sub-nebula model for the CPD to reproduce the Galilean satellites. Their results have showed that the presence and location of the snowline could explain the different ice-rock compositions of the Jovian moons. \citet{cilibrasi18} used a hydrodynamical simulation on Jupiter's formation to serve as an input for the CPD's density- and temperature profiles. In this case, the authors also included mass influx from the circumstellar disc  \citep{Szulagyi14} and a realistic dust coagulation model was used to determine the dust density profile \citep{Drazkowska18}, where the satellite seeds were forming via streaming instability. While \citet{Shibaike17} pointed out that the dust drift is an issue to keep solids and create satellitesimals, \citet{Drazkowska18} showed that dust traps and the outflowing gas stream can mitigate this effect and therefore satellitesimal formation is not only possible, but it happens very efficiently due to the short orbital time-scales around the planet. \citet{Szulagyi18} used hydrodynamic simulations of CPDs of Uranus and Neptune, and population synthesis for their satellite formation to understand how those satellite systems could have formed. They found that the CPD that naturally forms during the formation of these ice giants can form Uranian-like moons systems. In the case of Neptune, such systems might have existed before Triton capture happened and wiped away the original satellite system. Recently, \citet{Shibaike19} suggested that pebble accretion can reproduce many characteristics of the Galilean satellite system. \citet{RonnetJohansen19} successfully explained the formation of satellite systems around Jupiter and Saturn via pebble accretion as well.
    Other past models did not rely on the population synthesis approach. For example, \citet{Fujii17} used a semi-analytical model centered around migration and very successfully reproduced the Galilean moons' 1:2:4 mean resonance. \citet{Canup06} used an N-body model to investigate the origin of the total mass of the satellite systems, which in our Solar System are very similar for all gas giants. \citet{OgiharaIda12} also used an N-body approach in order to replicate the configuration of the Galilean moons.
 
    Given the prospects of upcoming exomoon discoveries, here we study satellite formation with a population synthesis approach in GI generated CPDs from \citet{Szulagyi16b}. Given that the CPD characteristics, such as the disc radius as well as the temperature and mass influx from the circumstellar disc changes with semi-major axis of the planet, we rescaled our nominal CPD simulation at various orbital separations from the star. We compare the satellite systems with CA models as well as giving detectability predictions of the formed exomoons. 
	
	\section{Methods}
	
	In this work we perform population synthesis (Section \ref{sec:semi}), on a base of hydrodynamical simulation (Section \ref{sec:hydro}) obtained disc profiles.
	
	\subsection{Hydrodynamical simulations}
	\label{sec:hydro}
	
	For the circumplanetary disc density- and temperature profiles, we use a hydrodynamical simulation output from \citet{Szulagyi16b}. This is a 3D global disc Smoothed Particle Hydrodynamic (SPH) simulation to study planet formation via disc instability. It is designed to mimic the formation of the HR 8799 system, which has several gas giants on wide (R > 30 AU) orbits, that might have formed via gravitational instability. 
	
	The circumstellar disc radius initially is set to 5-200 AU and 0.6 $M_{\rm{Solar}}$ mass. The initial surface density profile is set up in such a way that it follows a power law, where the exponent is close to $-1$ in the region of about 30 - 100 AU. For the temperature profile, an iterative procedure is used, which includes full force balance and stellar irradiation at t = 0 as well as disc self-gravity \citep{Mayer16,Rogers11}. The stellar mass is 1.35 $M_{\rm{Solar}}$. There are several clumps that form within the circumstellar disc. We extract one of the smallest clumps, where the planet mass is $\sim 10 \text{ } M_{\rm{Jupiter}}$ with an equal mass CPD.
	
	The simulation is run with the ChaNGa Tree+SPH code and contains 42 million particles. The finest resolution is 0.01 AU. Monaghan viscosity is used with $\alpha = 1$ and $\beta = 2$ . The code uses radiative cooling, which is dependent on local gas properties, and gives an energy loss per time per volume like:
	
	\begin{equation}
	\label{eq:RadCooling} 
	\begin{aligned}
	\Lambda = (36 \pi)^{1/3} \frac{\sigma}{s}(T^4 - T_{\rm{min}}^4)\frac{\tau}{\tau^2 +1}
	\end{aligned}
	\end{equation}
	where $\sigma$ is the Stefan--Boltzmann constant, $s = (m/\rho)^{1/3}$, $T_{\rm{min}}$ is the minimum gas temperature background (around 10 K in this case) and $\tau$ the optical depth. This way, cooling is most efficient at around $\tau = 1$, while it also recovers the dependence on $\tau$ of the cooling rate in the asymptotic limits of optically thin and thick discs. It does not solve the radiation hydrodynamic equation and as such does not take into account the accretional luminosity of contracting clumps, but it includes compressional heating (which is generated by PdV work) and shock heating.
	
	Optical depths are computed using the tabulated values from \cite{Dalessio97} for the Rosseland mean opacity and from \cite{Dalessio01} for the Planck opacities, assuming solar metallicity and dust-to-gas ratio 0.01. To take into account the ortho/para ratio of molecular hydrogen, which varies as a function on temperature and is important to recover the thermodynamics across spiral shocks that occur in self-gravitating unstable discs \citep{Podolak11}, a variable adiabatic index is used.
	
	\subsection{Population Synthesis}
	\label{sec:semi}
	
	In this 1D semi-analytical model, the initial circumplanetary disc profile is taken from the hydrodynamic simulation of \citet{Szulagyi16b}. Within the disc, satellite seeds are created, which can migrate, accrete, get captured into resonances, collide and get lost into the planet. At the same time, the disc is evolved too, both in density and temperature. A fixed grid is used within the CPD, with a cell size of a Jupiter radius and a fixed timestep $dt = t_{\rm{disp}}/1000$, where 
	the dispersion time-scale $t_{\rm{disp}}$ is a typical time-scale on which the disc disperses. This parameter is set at the start of the simulation (see Section \ref{sec:popsynth}). The simulation ends once the system has reached a steady state, which is always before 14 $t_{\rm{disp}}$.
	
	\subsubsection{Disc structure and evolution}
	\label{sec:disk}
	
	\begin{figure}
		\centering
		\includegraphics[width=0.5\textwidth]{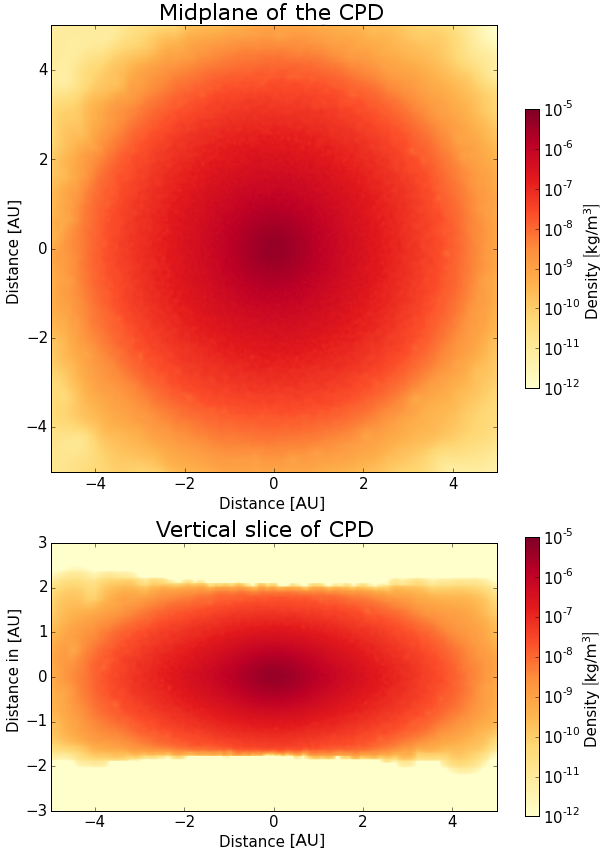}
		\caption{Density maps of the CPD midplane (top) and vertical slice (bottom).}
		\label{fig:DiskCuts}
	\end{figure}
	
\begin{figure}
		\centering
		\includegraphics[width=0.5\textwidth]{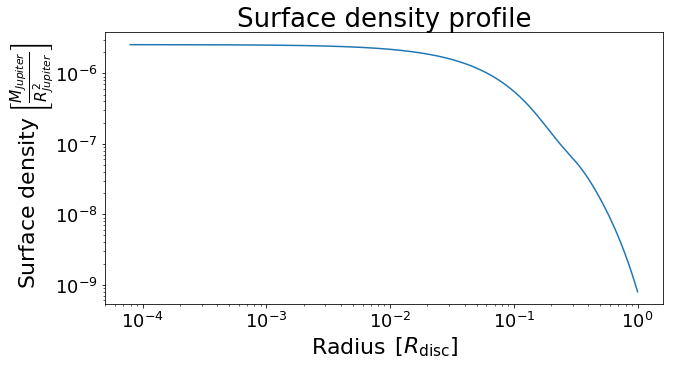}
		\caption{This plot shows the surface density as a function of disc radius.}
		\label{fig:surface}
	\end{figure}

\begin{figure}
		\centering
		\includegraphics[width=0.5\textwidth]{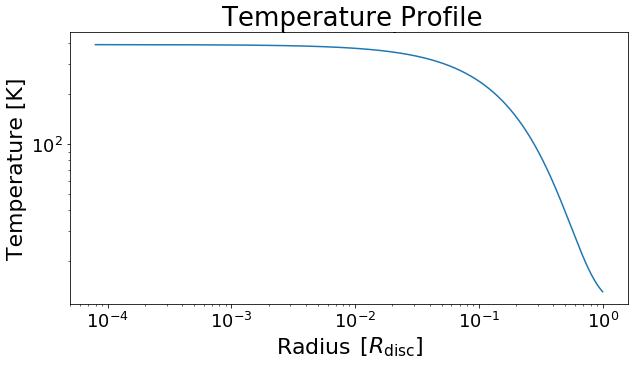}
		\caption{This plot shows the temperature profile as a function of disc radius.}
		\label{fig:temperature}
	\end{figure}

	Because the population synthesis is performed in 1D, we azimuthally average and vertically integrate the CPD from the 3D hydro simulation. The initial density profile is shown on Fig. \ref{fig:surface}, while the initial temperature profile is on Fig. \ref{fig:temperature}. Fig. \ref{fig:DiskCuts} shows a density map of the CPD midplane and a vertical slice.
	
	The CPD ranges from 3 $R_{\rm{Jupiter}}$ (as the planet is slightly bigger than 2 $R_{\rm{Jupiter}}$ and the grid size is taken to be 1 $R_{\rm{Jupiter}}$) to 12555 $R_{\rm{Jupiter}}$, i.e about 60 \% of the planet's Hill-radius or 6 AU (according to the hydro simulation, see Section \ref{sec:hydro}), and has a viscosity $\alpha = 0.004$, equal to the value in the hydrodynamic simulation. The code units are $R_{\rm{Jupiter}}$ for the radius, $M_{\rm{Jupiter}}$ for the mass, Kelvin for the temperature, and years for the time. The initial profiles for gas density and temperature are fits to the simulation data and take the following form
	
	\begin{equation}
	\label{eq:GasFit} 
	\begin{aligned}
	\Sigma_{\rm{gas,0}}(r) = 2.56 \cdot 10^{-6} \cdot e^{-2.57 \cdot \frac{r_{\rm{scaled}}}{1 \text{ AU}}} \\
	+ \frac{1}{ 1 + e^{-3.22 \cdot (\frac{r_{\rm{scaled}}}{1 \text{ AU}}-1.49)}} \\
	\times 3.02 \cdot 10^{-7} \cdot e^{-0.99 \cdot \frac{r_{\rm{scaled}}}{1 \text{ AU}}}  \text{ } \left[ \frac{M_{\rm{Jupiter}}}{R_{\rm{Jupiter}}^{2}}\right]
	\end{aligned}
	\end{equation}
	
	\begin{equation}
	\label{eq:TempFit}
	\begin{aligned}
	T_{\rm{0}}(r) = 381.14 \cdot e^{-0.87\frac{r_{\rm{scaled}}}{1 \text{ AU}}} \\
	+ (1-e^{-0.87 \cdot \frac{r_{\rm{scaled}}}{1 \text{ AU}}}) \cdot 11 \cdot \left(\frac{50 \text{ AU}}{a}\right) \text{ } [K]
	\end{aligned}
	\end{equation}
	
	The radius used here is adjusted relative to disc size, that is $r_{scaled} = r \cdot \frac{12555}{r_{out}}$ with $r_{out}$ the edge of the disc and 12555 is the outer edge in the reference cast of 50 AU), so that if the disc is smaller, the values for gas density are assigned based on relative position in the disc, while $a$ is the semi-major axis of the planet. See \ref{sec:diffa} for further explanations and the reasons for this scaling. The dust density is assumed to have the same shape as the gas density, but multiplied by a constant dust-to-gas ratio, which we change in the different 1D runs. The temperature, gas and dust densities also evolve with time. For both, an exponential decay \citep{Ida08} is assumed on a time-scale $t_{\rm{disp}}$:
	
	\begin{equation}
	\label{eq:dropG} \Sigma_{\rm{gas}}(r,t) = \Sigma_{\rm{gas,0}}(r) e^{\frac{-t}{t_{\rm{disp}}}}
	\end{equation}
	
	\begin{equation}
	\label{eq:dropT} T(r,t) = T_{\rm{min}} + (T_{\rm{0}}(r) - T_{\rm{min}})e^{\frac{-t}{t_{\rm{disp}}}}
	\end{equation}
	where $T_{\rm{min}}$ is the temperature of the gas surrounding the CPD, with a value of 11 $K$.
	
	While this CPD is relatively massive, it is not self-gravitating, because the Toomre Q parameter \citep{Toomre64} is always above unity. This is consistent with the absence of spiral structure and the large thickness of the CPD  (Figure \ref{fig:DiskCuts}).

	\subsubsection{Satellite formation and evolution}
	\label{sec:sats}
	
	\paragraph{Satellite formation}
	\label{sec:satform}
	
    In this model a similar approach is used as in \cite{Miguel16}, which means that embryos are created in the disc with a mass of $10^{-7} M_{\rm{Jupiter}}$. The first generation is created between $ [0,t_{\rm{disp}}/2] $, and on positions between $ 1\% -- 80\% $ of disc radius. 
    These positions are uniformly distributed between $[log_{10}(0.01 \cdot R_{\rm{disc}}),log_{10}(0.8 \cdot R_{\rm{disc}})]$. A log-uniform distribution is chosen because the	disc radius spans multiple orders of magnitude. We chose this large zone because the disc is massive enough that, even though dust densities become lower in the outer disc, the isolation mass (the total mass within a feeding radius of an object the mass of our seeds) is more than enough even at 0.8 $ R_{\rm{disc}}$ to form such a seed. Combined with the low temperatures, it should be possible for seeds to form there.
	
	The model also allows for embryos to be created at later times, in a sequential manner, to replace satellites that are lost into the planet \citep{Canup02}. Using the results from \cite{Shibaike17} to estimate the growth time-scale on which these later generation seeds form, which for our disc and in the range where we create embryos is between $[10^{4},10^{5}]$, which is also exponentially suppressed on the dispersion time-scale to reflect the decay in dust density. If an embryo is created, we randomly choose a time in said range after which a new embryo will be created. We choose this growth time randomly to reflect that, in reality, as these embryos grow to our threshold mass of $10^{-7} M_{\rm{Jupiter}}$, they would move themselves and thus 2 embryos that are inserted at the same positions can still have different growth time-scales. It should also be stressed that this growth time-scale of embryos (that is the time between the insertion of 2 embryos) is very different from the growth time-scale of satellites (that is the time-scale that is governed by accretion of dust as given in Sect. \ref{sec:accretion}).

	\paragraph{Migration}
	\label{sec:mig}
	
	Both Type I and Type II migration are considered and the two regimes are separated by gap opening. At first, when the satellites are not too massive, Type I migration will occur. As we assume circular and co-planar orbits, migration manifests as a change in the orbital radius $a$ and the migration prescription given by \cite{Tanaka02} is used:
	
	\begin{equation}
	\label{eq:Tanaka02} v = \frac{da}{dt} = b_{\rm{I}}\frac{ M_{\rm{sat}} \Sigma_{\rm{gas}} a^3 }{M_{\rm{planet}}^2} \left (\frac{a}{h} \right)^2 \Omega_{\rm{K}}
	\end{equation}
	where  $h = \frac{c_{\rm{s}}}{\Omega_{\rm{k}}}$ is the scale-height of the disc, $\Omega_{\rm{K}}$ is the Keplerian velocity at radius $a$, $c_{\rm{s}}$ is the local speed of sound, $M_{\rm{sat}}$ is the satellite mass and $M_{\rm{planet}}$ is the planet mass. The factor $b_{\rm{I}}$ is a correction
	factor that depends on the disc for which the prescription by \citep{Pardekooper10,Pardekooper11} is used in this work. The influence of the satellite on the gas density around it is also taken into consideration with a partial gap model. This is done by multiplying the velocity with a factor between 0 and 1, representing this depth, based on an analytical model by \citet{Crida07}.
	
	As satellites increase in mass, they might entirely open a gap in the disc. The following gap opening condition is used \citep{Crida06}:
	
	\begin{equation}
	\label{eq:P} P = \frac{3}{4}\cdot \frac{h}{R_{\rm{H}}} + \frac{50}{q\cdot Re} = \frac{3}{4} \frac{c_{\rm{s}}}{\Omega_{\rm{K}} a} \left(\frac{q}{3} \right)^{-1/3} + 50 \alpha q \left(\frac{c_{\rm{s}}}{\Omega_{\rm{K}} a} \right)^{2} > 1
	\end{equation}
	where $R_{\rm{H}} = a\cdot \sqrt[\leftroot{-1}\uproot{2}\scriptstyle 3]{\frac{q}{3}}$ is the satellite's Hill-radius, $c_{\rm{s}}$ the local speed of sound,
	$q = \frac{M_{\rm{sat}}}{M_{\rm{planet}}}$ and $Re$ is the Reynolds number. As the gap opening in particular is very dependent on the choice of the scale-height $h = \frac{c_{\rm{s}}}{\Omega_{\rm{k}}}$, we can justify this prescription by recognizing that the disc is not self-gravitating (see Sect. \ref{sec:disk}) and we can assume that most of the disc mass will be concentrated around the midplane. However, recent works 
	\citep{Malik15,Mueller18} have shown that for massive discs, such as the one assumed here, the above criterion leads to easier gap opening compared to the results of hydrodynamical simulations. Moreover, 
	\citet{Mueller18} showed that one can better match the gap opening conditions in simulations by additionally requiring that the satellite's crossing time, i.e. the time it 
	would take the satellite to migrate across a region with the width of the gap, $\tau_{\rm{cross}} = R_{\rm{HS}}(\frac{da}{dt})^{-1}$ \citep{Malik15} with $R_{\rm{HS}} = 2.5\cdot R_{\rm{H}}$ \citep{Pardekooper09}, is longer than the viscous time of the gas ($\tau_{\rm{visc}} = \frac{a^2}{\nu}$), so Type II
	migration happens if:
	
	\begin{equation}
	\label{eq:type2final}  P < 1 \text{ and } \tau_{\rm{visc}} < \tau_{\rm{cross}}
	\end{equation}
	Once a gap is opened, the satellites migrate with the gap on the viscous time-scale given by:
	
	\begin{equation}
	\label{eq:type2} v_{\rm{r}}  = -b_{\rm{II}}\frac{\alpha c_{\rm{s}} h}{a} 
	\end{equation}
	The factor $b_{\rm{II}} = \frac{1}{1+\frac{M_{\rm{sat}}}{4\pi \cdot a \cdot \Sigma_{\rm{gas}}}}$ \citep{Syer95} reflects that the migration slows down the heavier the satellite is compared to the gap.
	
	\paragraph{Resonance trapping}
	\label{sec:resonance}
	
	Two satellites can get captured into resonances with each other, which tend to make them migrate together. This can
	be modeled such that as satellites move towards each other, there is a repelling force between the two. This will lead to an increase in the satellites' separation, which is modeled in our work like \citep{Ida10}:
	
	\begin{equation}
	\label{eq:resonance} \frac{\mathrm db_{\rm{i}}}{\mathrm d t} \simeq 7\left(\frac{\mid a_{\rm{i}} - a_{\rm{j}} \mid}{R_{\rm{H}}}\right)^{-4}\left(\frac{R_{\rm{H}}}{a_{\rm{i}}}\right)^{2}v_{\rm{K,i}}
	\end{equation} 
	with $R_{\rm{H}} = \left(\frac{m_{\rm{i}} + m_{\rm{j}}}{3M_{\rm{planet}}}\right)^{1/3} $, where $M_{\rm{planet}}$ refers to the planet around which the satellites orbit, is the Hill-sphere of their combined system. The inner satellite will feel half of this separation increase towards the central planet, while the outer one will feel the other half outwards. Resonant configurations could then be formed by a balance between inward migration due to disc-satellite interaction and outward migration due to satellite-satellite interaction.
	
	\paragraph{Collisions}
	\label{sec:collision}
	
	One of the restrictions of 1D models is that the collisions are trivial. To avoid an over-estimation of the satellite collisions (and therefore the final moon masses), we use a 2D approach when we test for collisions. Once a satellite is created, a random angle between $[0,2\pi]$ is assigned. This is updated with every timestep using $\Omega_{\rm{K}}$ and the test for collisions uses the actual 2D distance between two satellites. Hence, a collision happens only if this distance is smaller than the radius of the combined Hill-sphere of the two moons. When a collision happens in our model, we assume only the more massive one survives, and we add the mass of smaller body to the larger one. As this is a 2D model, we do not take into account close-encounters, as we can not appropriately model them.
	
	\paragraph{Accretion}
	\label{sec:accretion}
	
	The satellitesimals accrete mass while migrating. The solid-accretion model by \cite{greenberg91} is used:
	
	\begin{equation}
	\label{eq:accretion1} \frac{dM_{\rm{sat}}}{dt} = 2\sqrt{\frac{R_{\rm{sat}}}{a}}\Sigma_{\rm{dust}}a^2\sqrt{\frac{M_{\rm{sat}}}{M_{\rm{planet}}}}\Omega_{\rm{K}}
	\end{equation}
	where $R_{\rm{sat}}$ is the satellite's radius. The $\Sigma_{\rm{dust}}$ in this case is the average dust density over the entire feeding zone of the moons. \cite{greenberg91} gave a value for the radius of this zone of $R_{\rm{feeding}} = 2.3 \cdot R_{\rm{H}}$. This accretion is again multiplied by the same factor for the partial gap as the migration, because the dust and gas are assumed to be well coupled.
	
	\paragraph{Dust depletion and refilling}
	\label{sec:refilling}
	
	As the satellites are growing, the accreted mass is subtracted from the disc dust density, so the mass $M_{\rm{i}}$ in a given cell $i$ decreases in the following manner:
	
	\begin{equation}
	\label{eq:depletion} \Delta M_{\rm{i}} = \frac{M_{\rm{i}} \cdot M_{\rm{accretion}}}{M_{\rm{feeding zone}}}
	\end{equation}
	
	Like in the case of core-accretion simulations, the CPD is continuously fed from the circumstellar disc \citep{Szulagyi14}, even in these GI simulations. We calculate the net mass influx from the SPH simulation: $\dot{M}_{\rm{in,0}} = 7.44 \cdot 10^{-5} \frac{M_{\rm{Jupiter}}}{\rm{year}}$ at t=0. Since the circumstellar disc will eventually dissipate, the feeding will decrease in the same way. Therefore, the feeding rate is also decaying exponentially on the disc dispersion time-scale: $\dot{M}_{\rm{in}} = \dot{M}_{\rm{in,0}}e^{\frac{-t}{t_{\rm{disp}}}}$. This influx has the same dust-to-gas ratio as the disc, in order to be consistent. As this is the total influx into the disc, a model is needed to determine how the disc reacts to changes in the dust density, specifically how it refills the dust gaps left by the accreting satellites. For that, a model like \cite{cilibrasi18} is used. It is assumed that the disc is in equilibrium at the beginning and that it wants to return to that equilibrium on a certain time-scale $t_{\rm{refilling}}$ according to the following formula:
	
	\begin{equation}
	\label{eq:dust_refilling} \Delta \Sigma_{\rm{dust}} =  \frac{\Sigma_{\rm{dust,0}} - \Sigma_{\rm{dust}}}{t_{\rm{refilling}}} dt
	\end{equation}
	This refilling can only ever be as much as the influx $\dot{M}_{\rm{in}}$.
	
	\subsubsection{Varying Initial Parameters}
	\label{sec:popsynth}
	
	Some of the initial conditions are a priori unknown, such as the dust-to-gas ratio. The following parameters are varied randomly in the different population synthesis runs:
	
	\begin{enumerate}
		\item The dust-to-gas ratio of the disc is varied in the range $[10^{-3},10^{-1}]$, based on observational constraints of circumstellar discs \citep{Ansdell16}.
		\item The dispersion time-scale of the disc is in the range $[10^{4},10^{5}]$ years.
		The time-scales are intentionally short because gravitationally unstable CSDs
		undergo strong viscous transport as a result of gravitoturbulence \citep{ Durisen07}, hence they
		are expected to dissipate faster than the typical Myr time-scale inferred for low mass T Tauri disc samples (see e.g. \cite{Mamajek09}).
		\item The refilling time-scale of the disc is in the range $[10^{2},10^{5}]$ years. This is the widest range possible (where it is still always bigger than the timestep), as this process is the least known.
		\item The positions in which the initial embryos are created are randomized between 1\% and 80\% of the disc radius.
		\item The amount of initial satellite seeds are set between 5 and 20. We choose a range of numbers to get both situations where there are few seeding positions, but where satellites are created very fast as well as the reverse. It was tested that adding more seeds is not influencing the results.
	\end{enumerate}
	
	The dust-to-gas ratios, refilling time-scales and initial satellite positions are distributed according to a log-uniform distribution, while the dispersion time-scales are distributed exponentially \citep{Fedele10}. Instead of randomizing all the parameters, one can also choose to set a fix value for a certain parameter. By doing this for different values of the fixed parameter, it is possible to investigate the parameter's influence on the results. This is done for the dust-to-gas ratio, and for the dispersion- and refilling time-scales.
	
	\subsection{CPDs at different distances from the star}
	\label{sec:diffa}
	
	We also investigate the effects of the planet's semi-major axis, i.e. the CPD distances from the star. Other than the 50 AU nominal case, we consider 5, 20, 35 and 70 AU semi-major axes. For this purpose, we rescale the CPD properties appropriately:
	
	\begin{enumerate}
		\item Because the CPD size depends on the planet's Hill-radius \citep{Quillen98} -- which linearly scales with the semi-major axis -- the CPDs are scaled accordingly with $ \frac{50 \rm{AU}}{a}$, where $a$ is the planet's distance from the star.
		\item It is assumed that the CPD density profile is only dependent on the process that creates the planet (in this case gravitational instability) and the size of the disc (where the semi-major axis of the planet is relevant). The surface density profile is also assumed to be roughly the same regardless of the semi-major axis of the planet. This means that the 50 AU disc is mapped in such a way that the inner and outer edge match and the in between values are distributed relatively to the radii.
		\item For the CPD temperature, a simple scaling is assumed, based on the distance of the planet from the star: $a^{-1}$.
		\item The mass influx is also changed. Results from hydrodynamical simulations in \cite{Szulagyi17} showed that the influx decreases by a factor of 2 if the planet is moved 10 times farther out (so in this case the influx at 5 AU is double that of the influx at 50 AU), so the influx values are scaled according to $\dot{M}_{\rm{in}}(a,t) = 3.25  \cdot (\frac{a}{\text{AU}})^{-0.3} \cdot \dot{M}_{\rm{in}}(50\text{ AU},t)$, where $a$ is the semi-major axis of the planet.
	\end{enumerate}
	
	\subsection{Observational predictions}
	
	We also want to make a prediction on the observability of the satellites that are created in our population synthesis. For this, exoplanets that have already been discovered are considered, the satellites created in the population synthesis are placed in orbit around them (adjusted relative to the exoplanet size), and a detection probability is computed for every satellite, and averaged over all satellite systems. This way, it is possible to get an average probability per planet of detecting at least one exomoon.
	First, the detection probability for a given satellite around an exoplanet that orbits a star has to be determined. It has to be checked whether a given satellite is big enough compared to the star for modern instruments to be able to detect them via transits. The measure for this is called the transit depth, which is given as $\left( \frac{R_{\rm{sat}}}{R_{\rm{star}}} \right)^2 $. If this depth is bigger than the instrument threshold, we set a detection probability of 1, 0 else.
	Second, geometrical factors need to be accounted for. If the inclination of the exoplanet's orbit relative to the observer is high enough, there is a small portion of time where an orbiting satellite will pass above/below the star and we will not be able to detect it. The fraction of time a satellite spends in a region where we can not detect it (and thus the probability that we can detect it) can be calculated with the following formula:
	
	\begin{equation}
		\label{eq:inclinationCorrection} 1 - \frac{1}{\pi} \cdot arccos\left( \left( \frac{R_{\rm{star}}}{cos(i)} - a_{\rm{planet}} \right)/a_{\rm{sat}}  \right) 
	\end{equation}
	where $i$ is the exoplanet orbit inclination. This factor is used if $cos(i) >= \frac{R_{\rm{star}}}{a_{\rm{planet}} + a_{\rm{sat}}}$, otherwise it is simply 1. Another geometrical factor is the time a satellite will spend directly in front of or behind the exoplanet and this factor we calculate like:
	
	\begin{equation}
			\label{eq:FrontBehindCorrection} 1 - \frac{2}{\pi} \cdot arcsin\left(\frac{1}{sin(i)} \sqrt{\left(\frac{R_{\rm{planet}}}{a_{\rm{sat}}} \right)^2 - cos(i)^2}  \right) 
	\end{equation}
	and it is be used if $cos(i) <= \frac{R_{\rm{planet}}}{a_{\rm{sat}}}$, else it is 1.
	These factors will then give a probability $p_{\rm{j}}$ of detecting a satellite $j$ around a given exoplanet. We then compute this probability for every satellite in a given satellite system created by the population synthesis and calculate the probability of detecting at least one of these satellites around a given exoplanet:
	
	\begin{equation}
			\label{eq:AverageProb} P_{\rm{sat}} = 1 - \prod (1 - p_{\rm{j}})
	\end{equation}
	
	This is done for every satellite system for a given exoplanet and averaged over the total amount of satellite systems to get the probability of detecting at least one satellite around a given exoplanet and then for every exoplanet in the catalogue (where the necessary information is found) to get in the end an average probability of detecting at least 1 satellite around an exoplanet.
	Since the exoplanets considered have varying masses and orbital radii, the satellite properties are rescaled to better fit around those planets.
	\begin{enumerate}
		\item Depending on the exoplanet's orbital radius, the satellite systems from population synthesis run with the CPD at a specific semi-major axis are considered. So if the exoplanet has an orbital radius $< 13$ AU, the satellites from the 5 AU case are considered, if the exoplanet is in the range 13 AU $< a_{\rm{exoplanet}} < 28$ AU the results from the 20 AU case are considered etc.
		\item The satellites' orbital radii are then rescaled as follows:
		\begin{equation}
				\label{eq:aSatRescale} a_{\rm{sat,\text{ } rescaled}} = \frac{a_{\rm{exoplanet}} - R_{\rm{exoplanet}}}{a_{\rm{CPD}} - R_{\rm{CPD \text{ } planet}}} \cdot a_{\rm{sat,\text{ }original}} + B
		\end{equation}
		where $R_{\rm{CPD \textbf{ } planet}}$ is the radius of the planet around which the CPD was formed and B is a constant defined such that $a_{\rm{sat,\text{ }rescaled}} = R_{\rm{exoplanet}}$ if $a_{\rm{sat,\text{ }original}} = R_{\rm{CPD\text{ }planet}}$.
		\item The satellite masses are assumed to be directly dependent on planet mass, so that the masses can be rescaled like
		\begin{equation}
				\label{eq:mSatRescale} m_{\rm{sat,\text{ }rescaled}} = m_{\rm{sat,\text{ }original}} \cdot \left(\frac{m_{\rm{exoplanet}}}{m_{\rm{CPD \text{ } planet}}} \right)^{\frac{1}{3}} 
		\end{equation}
		where $m_{\rm{CPD \text{ } planet}} = 10 \text{ } M_{\rm{Jupiter}}$ is the mass of the planet around which the satellites from the  population synthesis are created.
	\end{enumerate}

	\section{Results}
	\label{sec:results}
	
	We present our results in the following way. First we analyse a reference case,
	that of a CPD at 50 AU from the central star, describing the properties of the emerging satellites' population, and
	how those vary as we vary the key parameters such as dust-to-gas ratio, disc
	dispersion time-scale and dust refilling time-scale. Second, we compare the same
	properties for CPDs forming around disc instability proto-planets located at 
	different distances from the central star.
	
	\subsection{The reference case: a CPD at 50 AU}
	\label{sec:results1}
	
	\subsubsection{Number of satellites}
	\label{sec:NumSats}
	
	\begin{figure}
		\centering
		\includegraphics[width=0.5\textwidth]{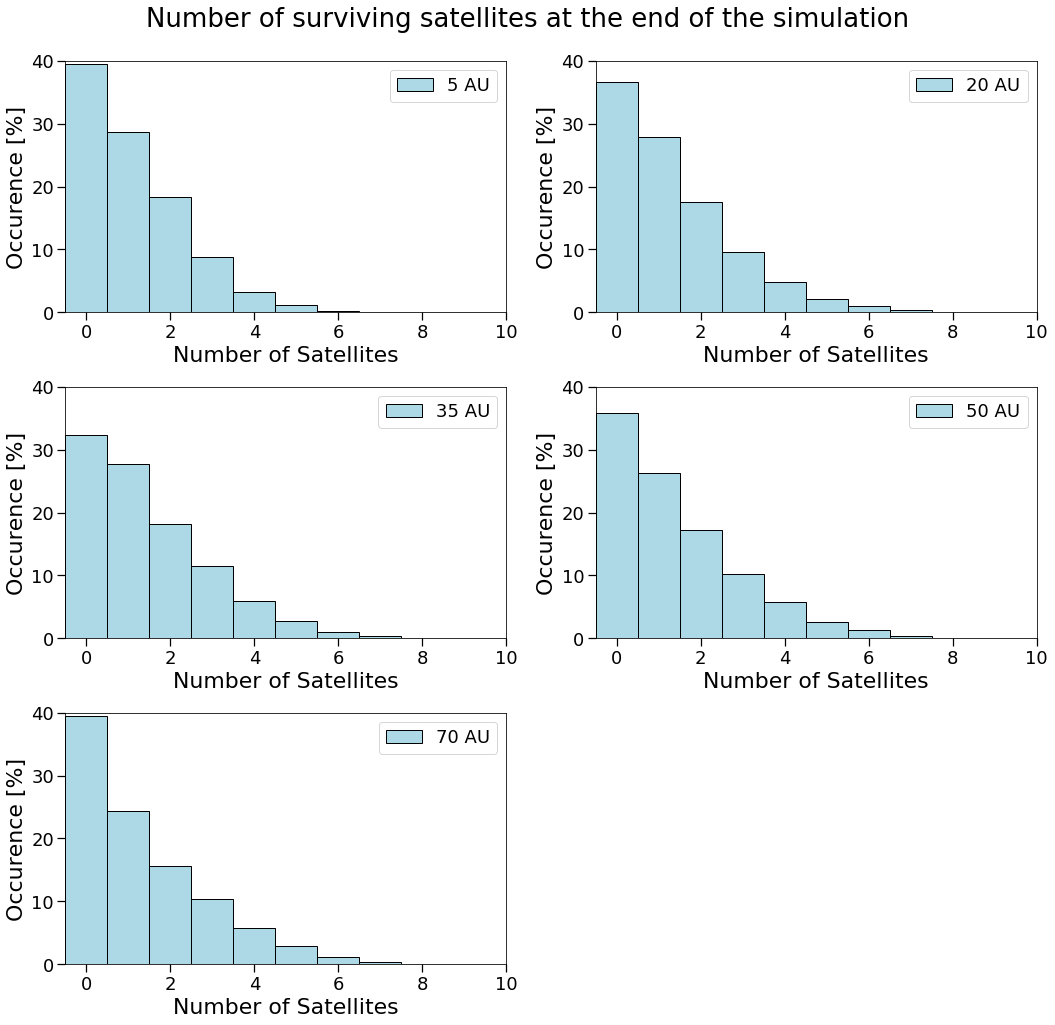}
		\caption{This plot shows a comparison of how many satellites with a mass greater than that $10^{-5}$ $M_{\rm{planet}}$ are surviving for the different 
		radial distances of the CPD from the star.}
		\label{fig:NumSats1}
	\end{figure}
	
	Fig. \ref{fig:NumSats1} shows the number of satellites that survive and are at least $10^{-5}$ planet masses. For the 50 AU case, in about 60 \% of cases, at least 1 satellite is created, with a maximum of 10 in less then 1 \% of the simulations. In the 35 \% of cases that no embryo reaches the cut-off mass, there are 0.045 \% cases where there is not even an embryo alive at the end of the simulation, which means they all either migrated into the planet or are lost to collisions.

\begin{figure}
		\centering
		\includegraphics[width=0.5\textwidth]{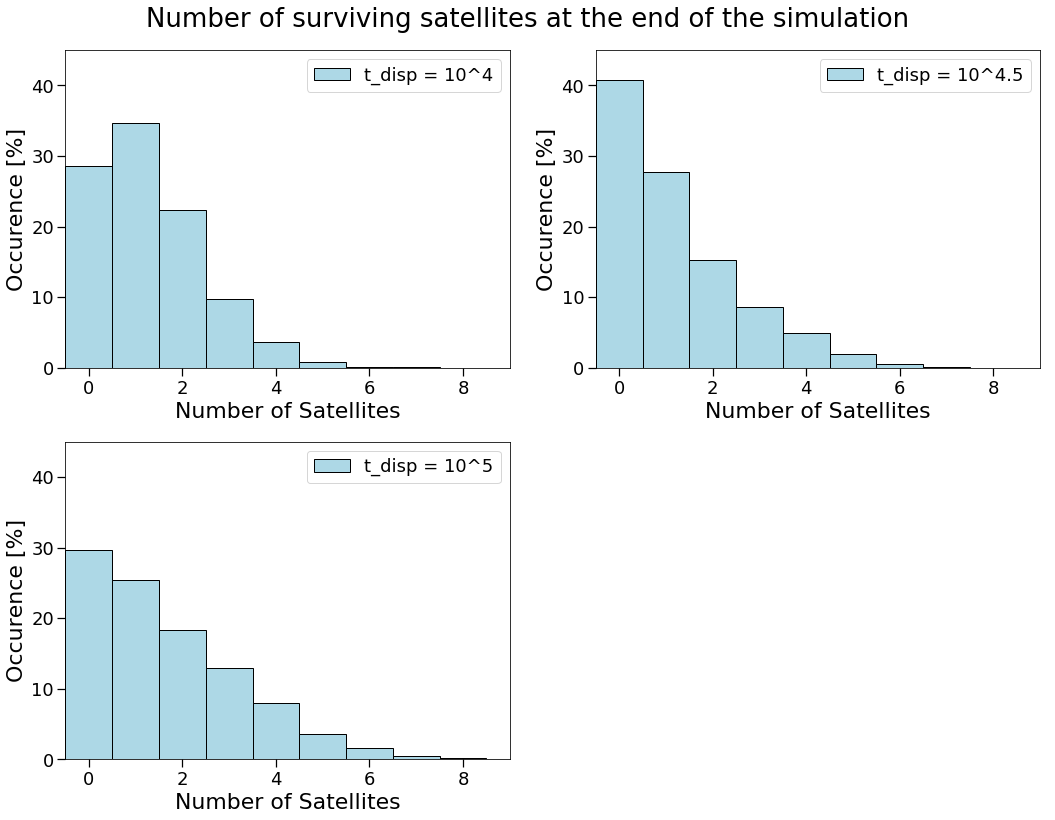}
		\caption{This plot shows in how many simulations a certain amount of satellites with a mass greater than $10^{-5}$ $M_{\rm{planet}}$ form in simulations with different, set values of the dispersion time-scale.}
		\label{fig:NumSatsDisp1}
	\end{figure}

	The dependence of the number of surviving satellites on dust-to-gas ratio and refilling time-scale is as expected. It reflects the fact that it is easier to reach a mass higher than $10^{-5}$ $M_{\rm{planet}}$ when there is more dust, or when the gap is refilled faster.
	Varying values for fixed dispersion time-scales give a non trivial picture though. Fig. \ref{fig:NumSatsDisp1} shows that as the dispersion time increases, the number of satellites first drops, then increases again. The reason for this behaviour is that the first generation of embryos will be the one that accretes the most mass, as the disc will have more dust available at earlier times. If the dispersion time-scale (and thus the disc lifetime) is low, less of those first generation embryos will eventually migrate into the planet. For intermediate dispersion time-scales, many first generation satellites will be lost into the planet, but subsequent generations will not have the time to grow enough to replace the population of high mass satellites. This can only happen in long dispersion time-scales, which is the behaviour Fig. \ref{fig:NumSatsDisp1} shows.

	\subsubsection{Mass distribution}
	\label{sec:SatsMass}
	
	\begin{figure}
		\centering
		\includegraphics[width=0.5\textwidth]{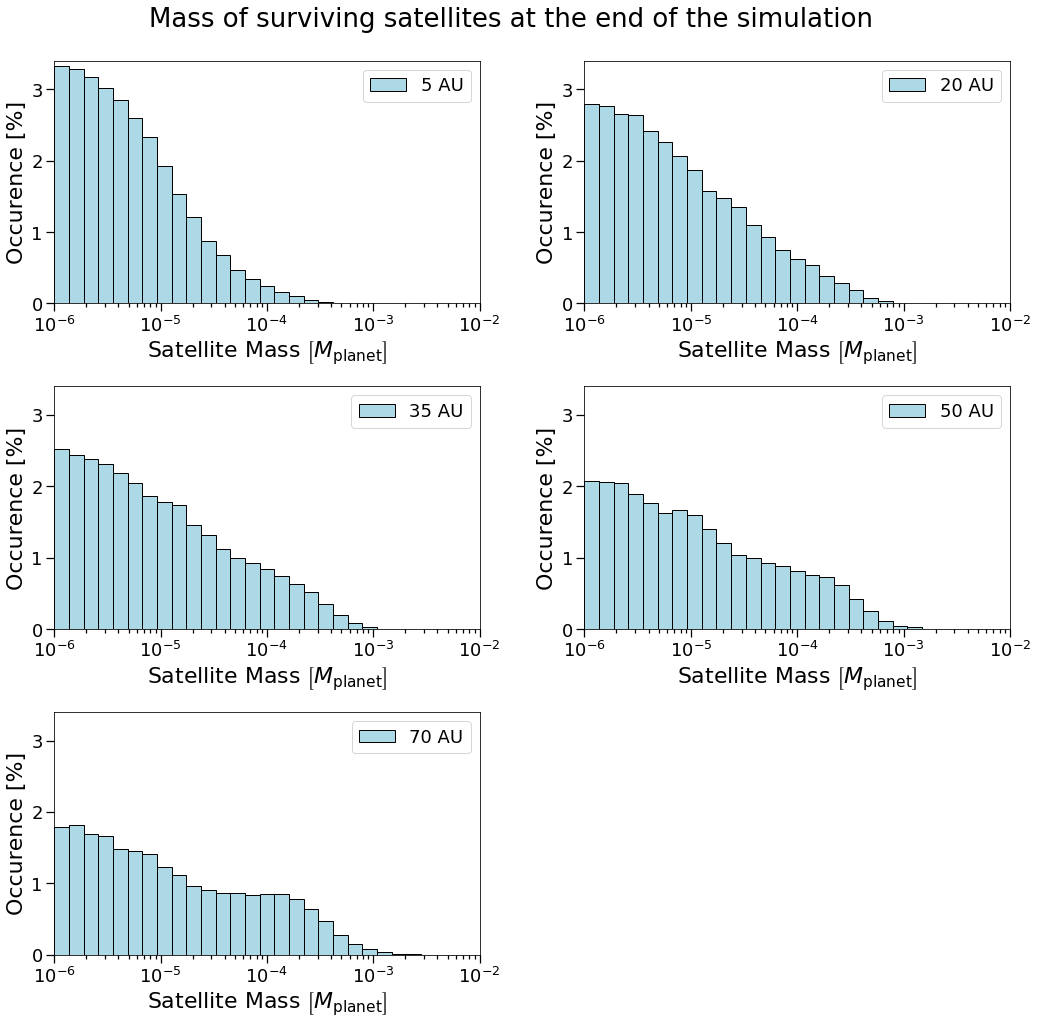}
		\caption{This plot shows a comparison of the masses of surviving satellites for the different central planet positions. The masses below $10^{-6} \text{ } M_{\rm{planet}}$ are cropped because of 
		the way we spawn embryos, there are a lot of them that get created late enough
		that they simply do not accrete mass anymore, so the numbers are heavily skewed towards them. We see that as the planet is further out (and thus has a bigger disc), the satellites become in general heavier.}
		\label{fig:MassDistri1}
	\end{figure}
	
\begin{figure}
		\centering
		\includegraphics[width=0.5\textwidth]{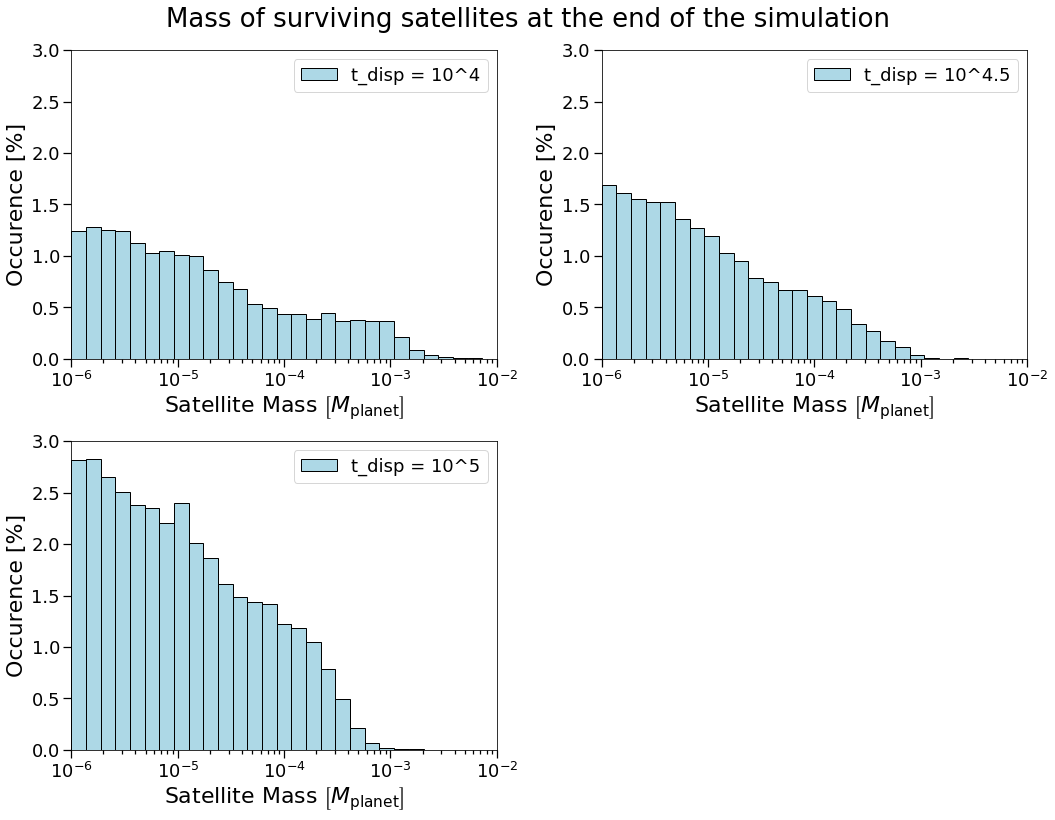}
		\caption{The distribution of the masses of all satellites that survive until the end of the simulation with a fix dispersion time-scale.}
		\label{fig:MassDistriDisp}
	\end{figure}

	Fig. \ref{fig:MassDistri1} shows the distribution of satellite masses as a fraction of the mass of the central planet, namely in units of 10 Jupiter masses. We see in the 50 AU case that many are in the range of Jovian satellites, between $10^{-6}$ and $10^{-5}$ planet masses, but they can reach as high as $10^{-3}$ planet masses, which corresponds to 3 Earth masses. This upper limit is a consequence of accretion slowing down and eventually stopping as dust is depleted from the disc by both accretion and dust dispersion as well as an increasingly smaller feeding radius due to migration towards the planet. Interestingly, such natural upper mass limit emerging from our model is of the order of the mass of observed Super-Earths, at which planets with H/He envelops are known to exist. This suggests that, in order to better follow the evolution of satellites in the upper tail of our mass function, including the possibility that they could accrete a gaseous envelope may be required. These masses are almost entirely accreted during migration as collisions are very rare.
    Varying values for dust-to-gas ratio and refilling time-scale yield a trivial outcome; higher dust-to-gas ratio implies higher masses, and shorter refilling time-scale implies more satellites can form at any given mass. The results for the dispersion time-scales are shown in Fig. \ref{fig:MassDistriDisp}. The behaviour corresponds to the result in Section \ref{sec:NumSats}; for fast dispersion times, there are many high mass satellites, close to $10^{-3}$ planet masses. These satellites vanish as the dispersion time-scale becomes higher, because they migrate into the planet. As the dispersion time-scale increases, there are more satellites in general, with a similar distribution, as now the subsequent generations have more time to accrete mass, but do so on a slow enough time-scale that they hardly migrate into the planet in the remaining available time.
	
	\begin{figure}
		\centering
		\includegraphics[width=0.5\textwidth]{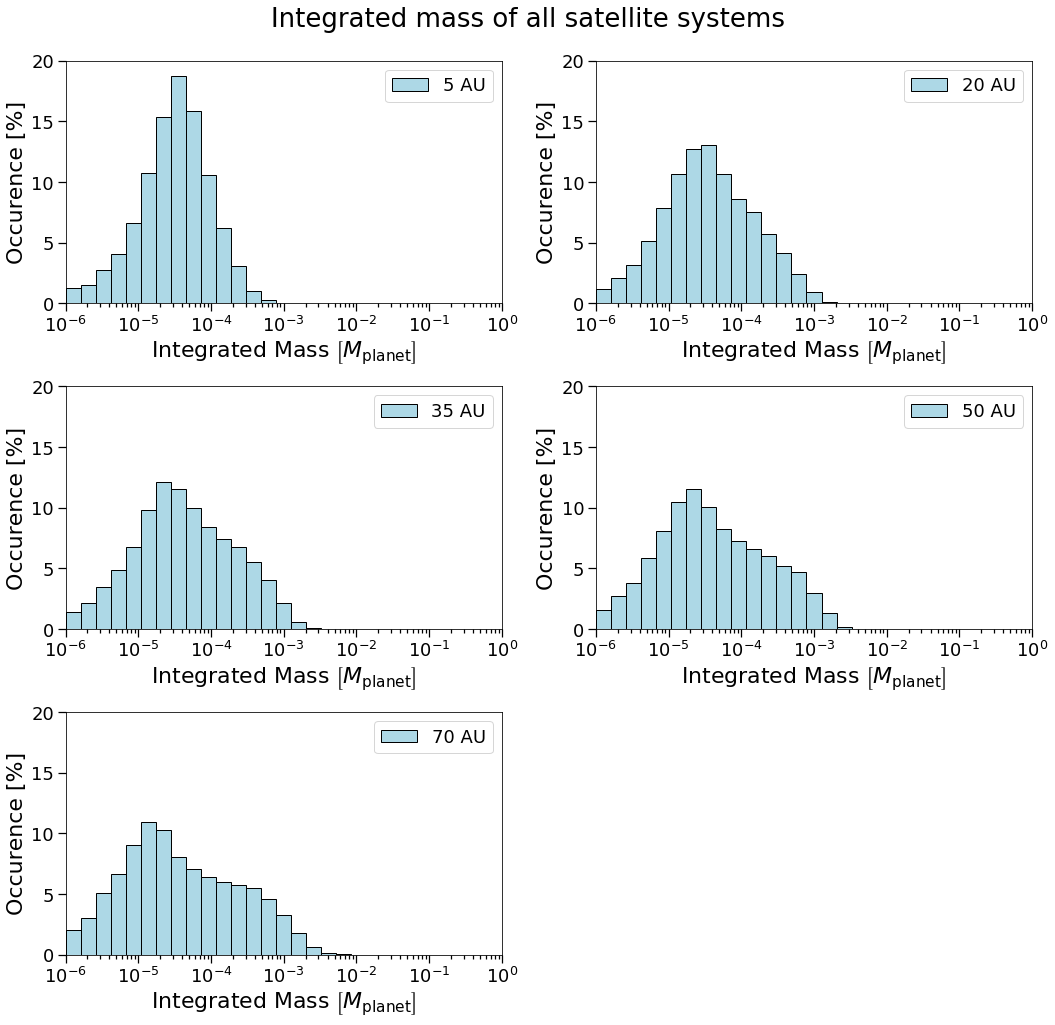}
		\caption{The integrated mass of the satellite systems for the different radial distances of the planet. They all have a maximum between $10^{-5}$ and $10^{-4}$ planet masses, but as the disc grows bigger the distribution grows wider.}
		\label{fig:IntMass1}
	\end{figure}
	
	Fig. \ref{fig:IntMass1} shows the integrated mass of the different satellite systems created. In our Solar System's gas giants, the satellite systems are consistently around $2 \cdot 10^{-4}$ planet masses. About 50 \% of the cases in the 50 AU case are between $10^{-5}$ and $10^{-4}$ planet masses, lower than what can be observed in the Solar System. About 30 \% are $10^{-4}$ planet masses and higher. As the maximum integrated mass is not much higher than the maximum mass possible for a single satellite, it also shows that satellites that massive are very rare and tend to dominate their systems.
	
	\begin{figure}
		\centering
		\includegraphics[width=0.5\textwidth]{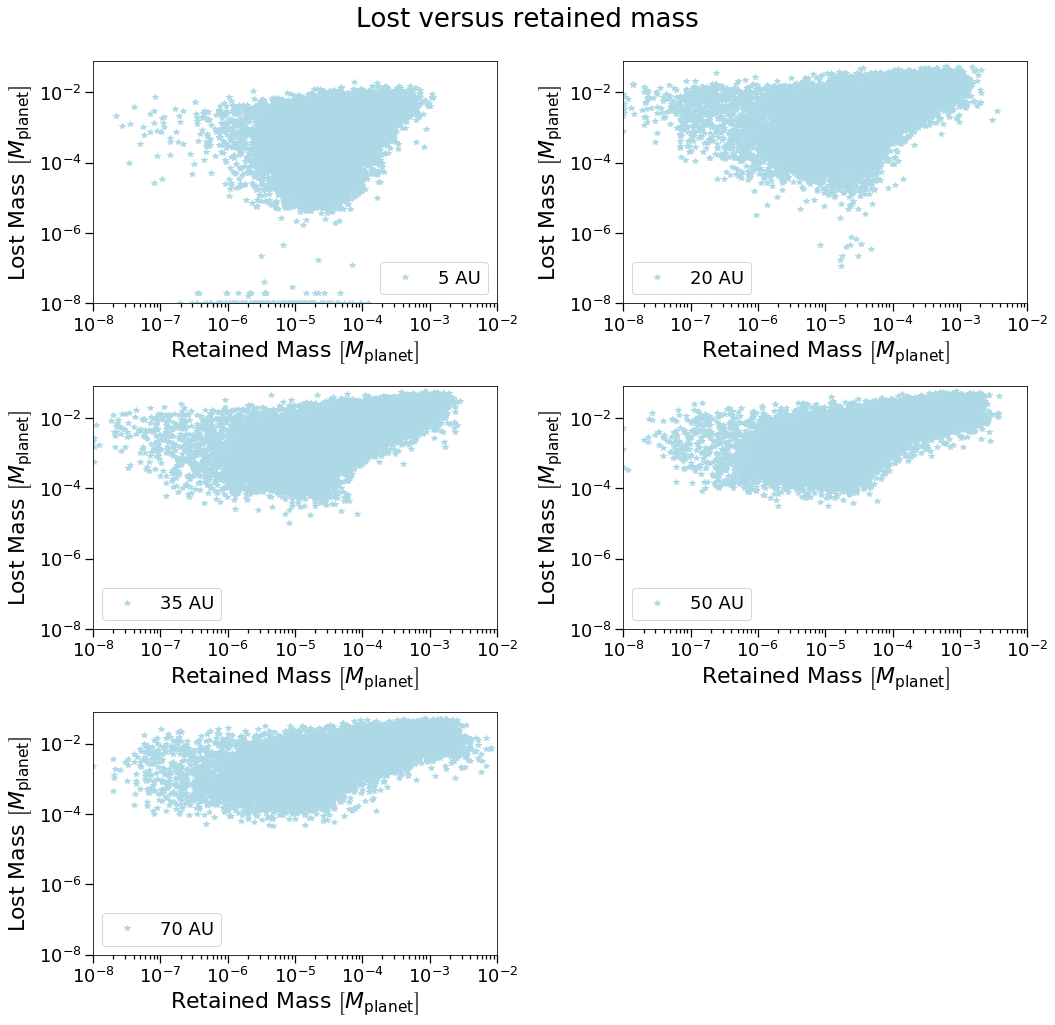}
		\caption{This plot shows the amount of mass lost into the central planet against the combined mass of surviving satellites. It shows that retained mass depends almost linearly on lost mass and that the retained mass is less or about the same as the lost mass.}
		\label{fig:LostMass1}
	\end{figure}
	
	Previous work \citep{Bolton17} has shown that Jupiter is about 20 times more
	enriched in metals relative to the solar metallicity value. Pollution by embryos migrating into the planet could explain this result. Fig. \ref{fig:LostMass1} displays the mass retained by the satellites versus the mass lost by them, whereby the median mass lost is about $3 \cdot 10^{-3}$ planet masses, or about 10 Earth masses. It shows that mass that is lost scales with the retained mass, where the latter is always lower than the former.
	
\begin{figure}
		\centering
		\includegraphics[width=0.5\textwidth]{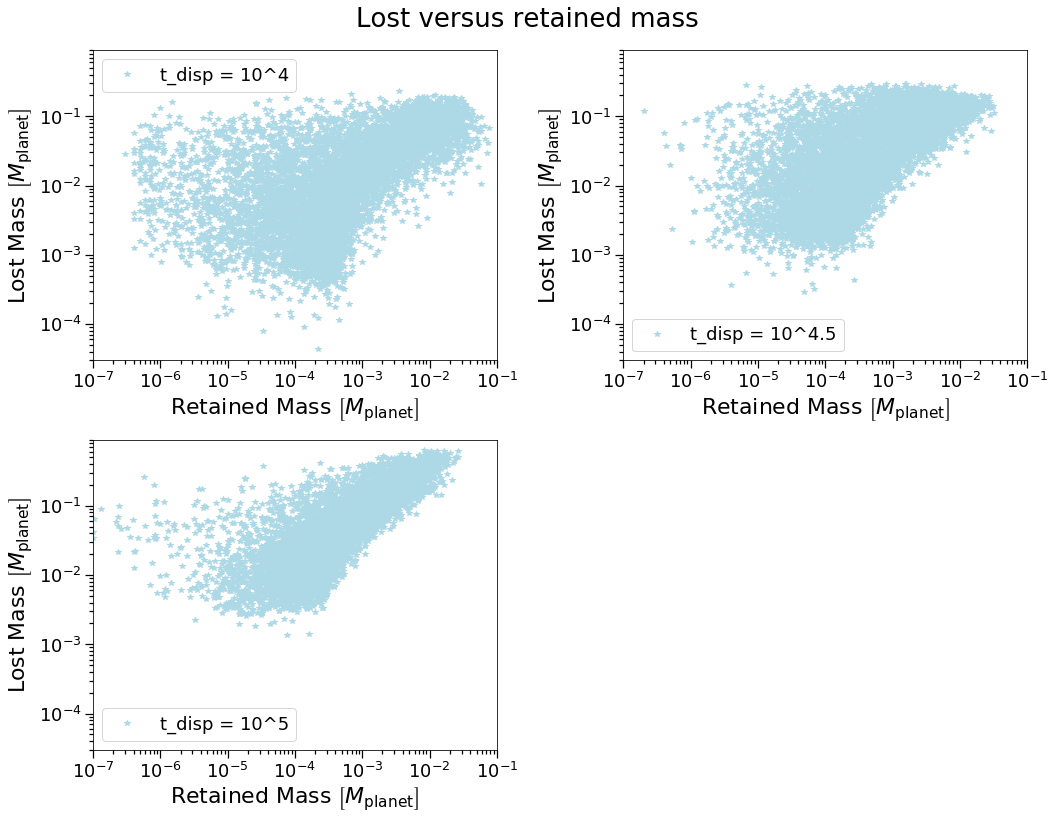}
		\caption{This plot shows the amount of mass lost into the central planet against the combined mass of surviving satellites from simulations with different, set values of the dispersion time-scale.}
		\label{fig:LostMassDisp}
	\end{figure}

	The results for varying values of fixed parameters mirrors those in Section \ref{sec:NumSats} and of the mass distribution: higher dust-to-gas ratio and faster refilling time-scales results in higher lost and retained masses. Fig. \ref{fig:LostMassDisp} shows that as dispersion time-scale increases so does the lost mass, which corresponds to the decrease in the number of heavy satellites which get lost into the planet. This also accounts for the decrease in the maximum retained mass. For long dispersion time-scales, both lost and retained mass increases, as then subsequent generations of satellites can accrete more mass.
	
	\subsubsection{Radial distribution of satellites}
	\label{sec:pos}
	
	\begin{figure}
		\centering
		\includegraphics[width=0.5\textwidth]{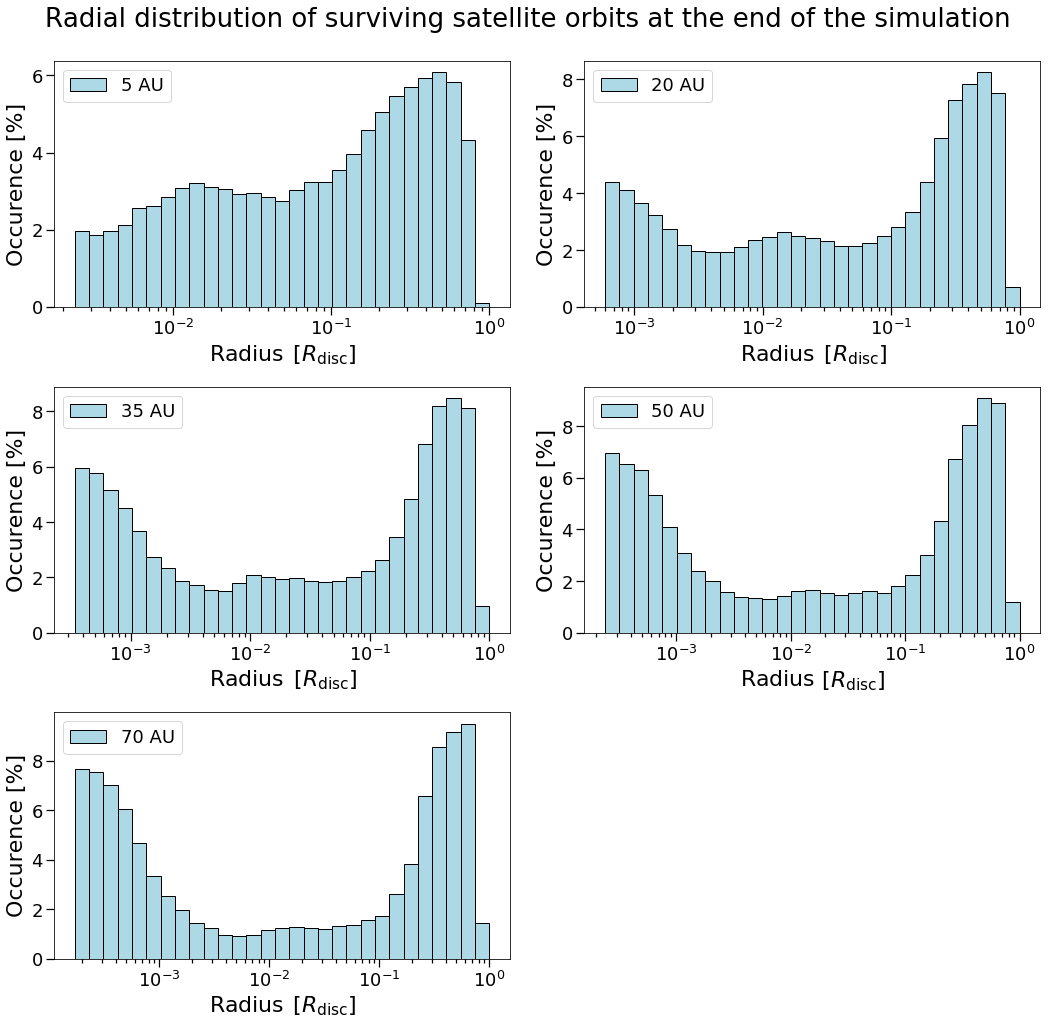}
		\caption{Comparison of satellite locations at the end of the simulations for the varied semi-major axis cases. When the planet is at 20 AU or further out, the histograms look roughly the same as the nominal (50 AU) case: one peak of satellites in the inner CPD and one peak in the outer disc. As the disc grows bigger, the inner peak grows as well, showing that as the disc size increases, it becomes less likely for satellites to migrate into the planet.}
		\label{fig:Pos1}
	\end{figure}
	
	Fig. \ref{fig:Pos1} shows the distribution of final positions for satellites with at least Europa mass. In the 50 AU case there is a peak close to the planet, where the semi-major axes of the orbits are comparable to the planet radius. Even though we account for resonant trapping, there are no resonant configurations that show up consistently, as they would manifest as separate peaks. As the distance from the planet increases, there are less satellites orbiting. Migration time-scales scale mainly with gas density and satellite mass and as the feeding zone will be small this close to the planet, satellites will have to accrete mass further out and migrate into this region. As gas density increases towards the planet, their migration will speed up, even though their mass will remain very stagnant and thus it is much more likely that any satellite entering the region past the close peak will migrate close to or into the planet eventually, leaving an evacuated area.
	
	It is more surprising that a second peak is located very far out. The expectation would be that a satellite which reaches that much mass would migrate away from its formation site so that the distribution should spread out more than what the plot shows. 
	The behaviour of satellites that are far out is investigated by specifically placing embryos in the region past the far peak and monitoring their evolution. These tests show that only embryos that accrete mass very fast in their early history can travel past the far peak. This is because migration speed is proportional to both gas density and the mass of the migrating object. The gas density (which in the region past the far peak is already very low) will drop over time because of the dispersion of the disc, but the mass will increase as the embryo accretes mass. For a satellite to be able to migrate past the far peak, it needs to overcome the low gas density and its drop over time with a fast enough accretion early in their life. This can only happen with certain combinations of dust-to-gas ratio, dispersion- and refilling time-scales. For most satellites, their accretion is too low to overcome the drop in gas density, resulting in very little migration. This means however that they can accrete mass throughout the whole disc lifetime, undergoing very few collisions with other proto-satellites, which makes it possible for them to reach the  mass of Europa  or higher. Therefore, the peak on the outer edge is a result of the interplay between satellite accretion and gas density.
	
	This second peak is not observed within our Solar System, which could be explained by the fact that the gas giants are all smaller and closer to the sun than our gas giant and also more likely to have been created by. This would lead to lighter CPDs with a smaller radius (see \ref{sec:discl} for a discussion on the difference of CPDs in the GI and CA case), which could inhibit the growth of far out embryos.
	
	We also analysed the changes in the results as we vary the main initial conditions separately. For higher dust-to-gas ratio, more satellites migrate close to the planet, because they grow more massive, which facilitates migration. The refilling time-scale has the same effect, although less pronounced. For short dispersion time-scales, there are also more satellites closer to the planet. These are the high mass satellites we discussed earlier, and as the dispersion time-scale increases, these will migrate into the planet, resulting in a shrinking peak in the histogram.
	
	\subsubsection{Formation temperature}
	\label{sec:formt}
	
	\begin{figure}
		\centering
		\includegraphics[width=0.5\textwidth]{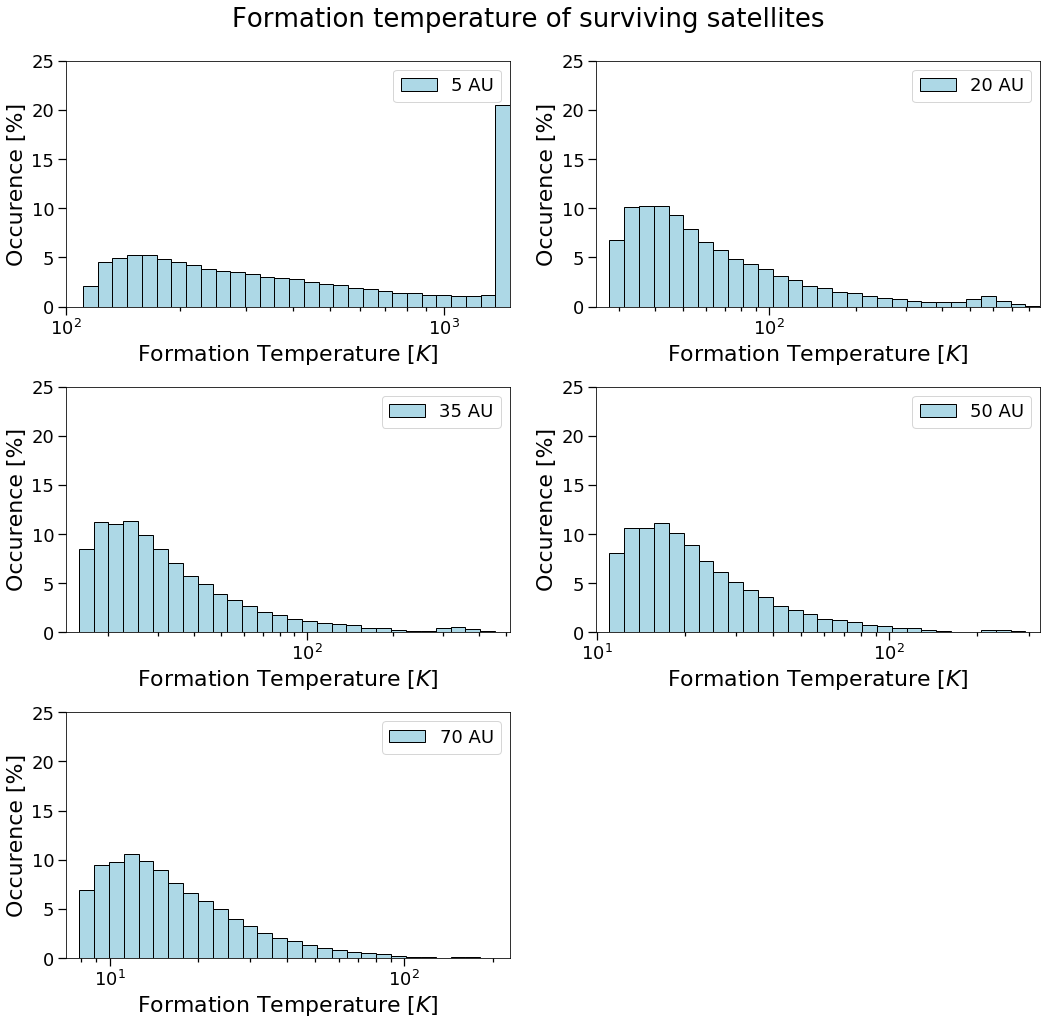}
		\caption{Comparison of formation temperature of the surviving satellites for the different cases. We see that even though the disc becomes hotter the closer the central planet is, most satellites will still be formed icy. For the 5 AU case, the peak at 1500 K is because at this temperature the dust evaporates, so there are no embryos formed at temperatures above.}
		\label{fig:Temp1}
	\end{figure}
	
	We also look at the temperature of the gas at the starting location of each satellite, as that should be indicative of the composition of the satellites. As water in the disc freezes at about 180 K \citep{Lodders03}, we can at least 
	use the formation temperature to estimate the likelihood to obtain icy or rocky satellites. Fig. \ref{fig:Temp1} shows that in the 50 AU case the vast majority of satellites form in regions with temperatures below 100 K, which is a strong indication that in discs like this, most satellites should end up being icy. This is not surprising, as Fig. \ref{fig:temperature} shows that, even at the beginning, most of the disc is near or below 180 K.
	
\begin{figure}
		\centering
		\includegraphics[width=0.5\textwidth]{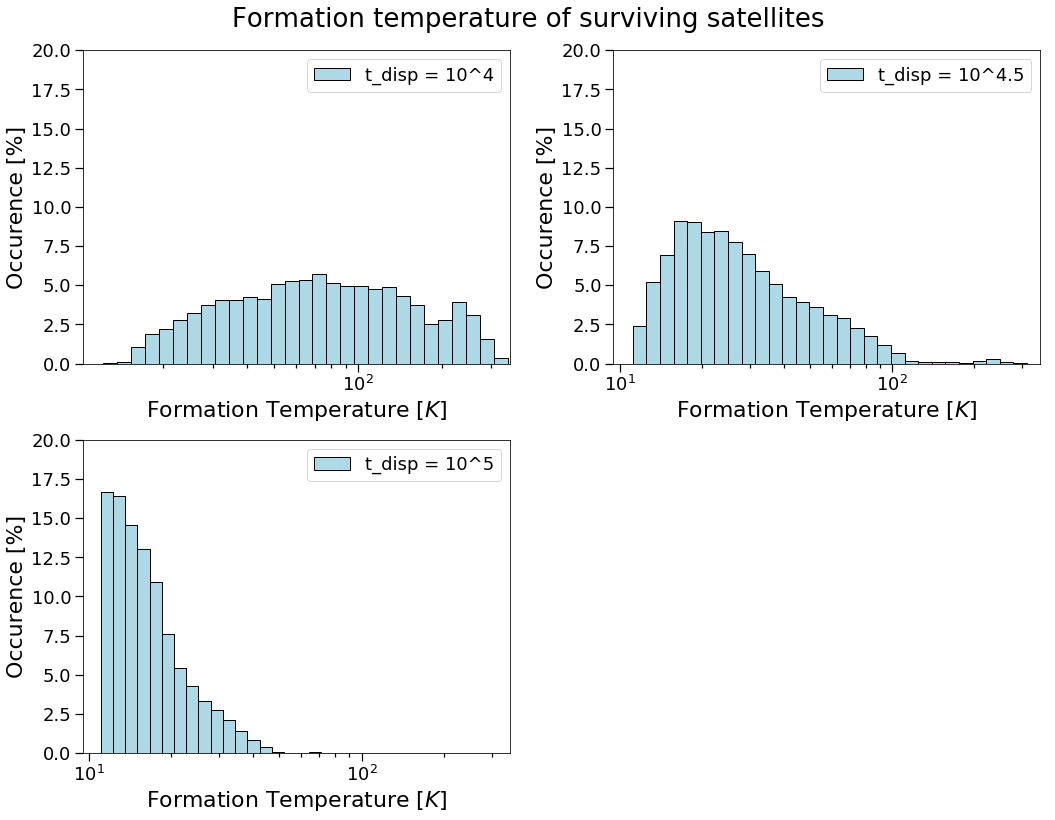}
		\caption{Formation temperature of the surviving satellites of at least Europa mass from simulations with a set dispersion time-scale.}
		\label{fig:TempDisp}
	\end{figure}

\begin{figure}
		\centering
		\includegraphics[width=0.5\textwidth]{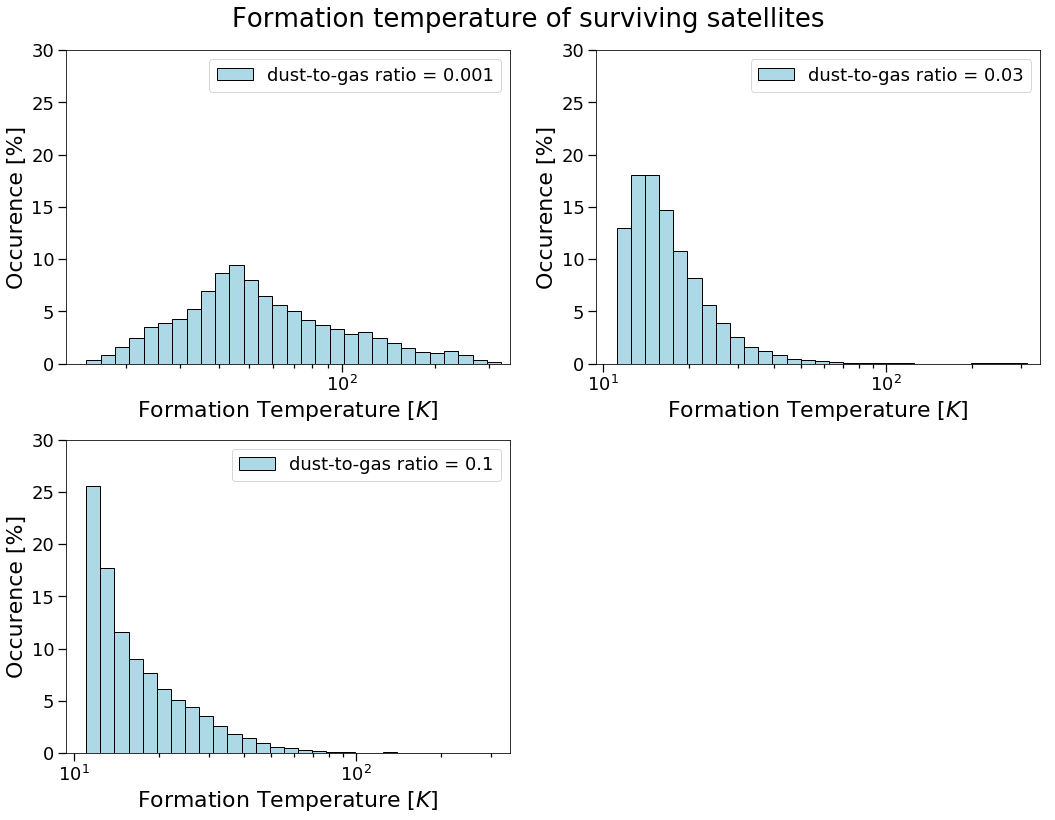}
		\caption{Formation temperature of the surviving satellites of at least Europa mass from simulations with a set dust-to-gas ratio.}
		\label{fig:TempDTG}
	\end{figure}

	We again checked the influence of varying the initial parameters one by one. The refilling time-scale has no apparent influence on the shape of the distribution. Fig. \ref{fig:TempDisp} shows the result for three different values for the dispersion time-scale. We see that for short dispersion time-scales, the formation temperatures are very high. That is because the short disc lifetime, which is coupled to the dispersion time, means that the surviving satellites will be older, as it is hard for them to migrate into the planet, and thus are created in an earlier, hotter disc.
	In Fig. \ref{fig:TempDTG} we investigated the effect of the dust-to-gas ratio. It shows that for low dust-to-gas ratios, the satellites will also be created in a hotter disc. That is because for this low amount of dust, the only embryos that will grow significantly in size, are necessarily older.
	
	\subsubsection{Time-scale to reach the Europa mass scale}
	\label{europa}
	
	\begin{figure}
		\centering
		\includegraphics[width=0.5\textwidth]{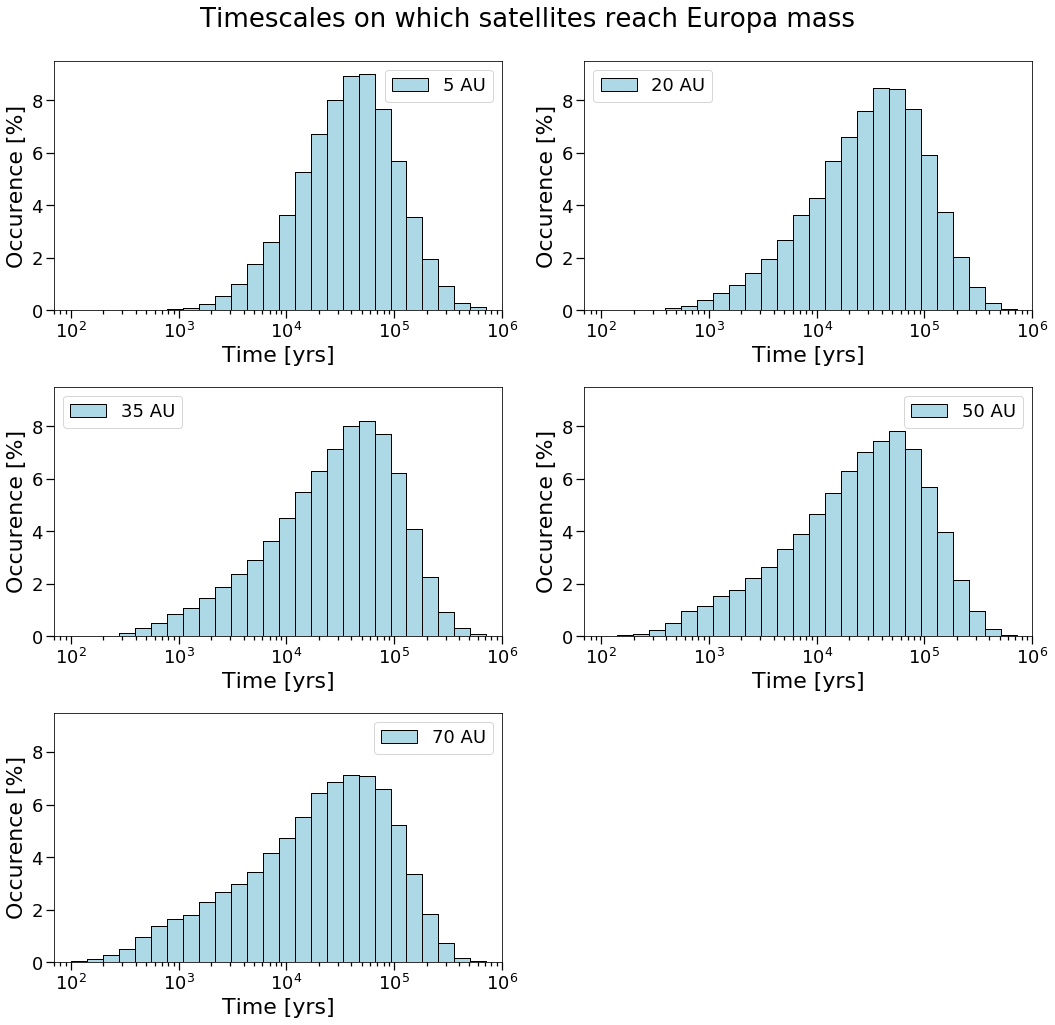}
		\caption{Comparison of time-scales to reach a mass equivalent to that of Europa in the different cases. Most of them peak around $10^{4}$ years, but as the disc grows bigger, the distribution widens. This is because in general the cells which make up the disc will contain more mass and it will thus be easier for satellites to accrete mass very fast.}
		\label{fig:Eu1}
	\end{figure}
	
	We also assess how fast the satellites form and use the time in which they reach Europa mass as a scale. In principle, their final composition and structure can be heavily influenced by how fast they form, and subsequently how much time they spend in the gas disc as satellites. We choose the mass of Europa as a reference because that corresponds  $10^{-6}$ the mass of the parent planet in our case, which turns out to be  the lowest mass significant satellites will have. Fig. \ref{fig:Eu1} shows that in the 50 AU case the majority of satellites will reach the Europa mass scale on a time-scale similar to the dispersion time-scale, between $10^4$ and $10^5$ years, which is quite long. The maximum is around $10^6$ years, which is just shorter than the maximum time a simulation can run in general ($14 \cdot 10^5$ years). However, it can also be very short, $10^3$ years and slower, although such cases are rare.
	
\begin{figure}
		\centering
		\includegraphics[width=0.5\textwidth]{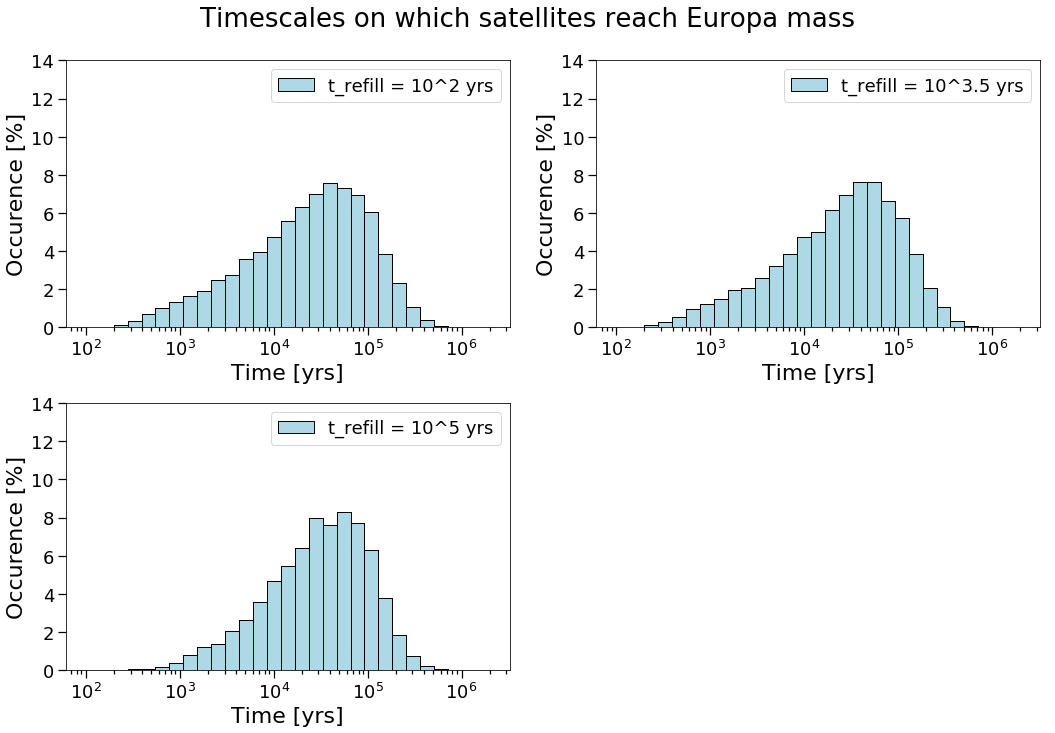}
		\caption{Comparison of Europa mass times in systems with a set refilling time-scale, but with the other parameters still randomized.}
		\label{fig:Eu3}
	\end{figure}

	We also consider the influence of the other other free parameters on the latter time-scale. Variations of the dust-to-gas ratio produce the expected result: the higher the ratio, the more satellites reach Europa mass very fast, which results in a much wider distribution. If we fix the dispersion time-scale, the distribution looks essentially the same as in the fully randomized case, but, since the disc lifetime is coupled to the dispersion time-scale, the bulk of the distribution shifts along with the dispersion time-scale. In Fig. \ref{fig:Eu3} we fixed the refilling time-scale at various values. It shows that the refilling time-scale only influences the very fast growing satellites, as a fast refilling means more satellites with a Europa mass time-scale in the $10^3$ years region. The rest of the distribution is largely the same, which means that the satellites with long Europa time-scales are satellites that migrate consistently enough, so that their mass comes from the disc itself, and not through influx into the gaps.
	
	\subsection{Comparison of CPDs at different distances from the star}
	\label{sec:results2}
	
	\subsubsection{Number of satellites}
	\label{sec:num2}

	Fig. \ref{fig:NumSats1} shows a comparison of how many satellites with masses higher than $10^{-5}$ planet masses survive until the end. The results are very similar, there is only a very slight increase in numbers as the disc grows in size. This is the result of two opposing effects: a bigger disc (which means a bigger semi-major axis of the planet) has more space for the satellites to transverse and more mass for satellites to accrete, so with increasing size the satellites should be more massive and thus more numerous. But as the distance from the star increases, the influx into the disc decreases, which means that embryos on average grow slower, which makes satellites less numerous. These two effects in this case result in the observed behaviour.
	We find that for the 5 AU and 70 AU cases, there is always at least one embryo that survives until the the end of the simulations, even if it doesn't reach $10^{-5}$ planet masses. For the 20 AU, 35 AU and 50 AU we find that there are 0.045 \%, 0.045\% and 0.02 \% cases respectively where no embryo survives until the end.

	\subsubsection{Mass distribution}
	\label{sec:mass2}
	
	Fig. \ref{fig:MassDistri1} shows the mass distributions. It shows that as the disc grows in size (and thus also in mass), the average mass of a satellite increases as well, as shown in the maximum masses reached. In the 5 AU case, satellites grow at most to about 1-2 Earth masses, while in the 70 AU case the maximum is at about 10 Earth masses. Secondly, it also shows that lower mass satellites grow relatively less likely, while higher mass satellites become more likely. In the 70 AU case, this even results in a slight peak close to the maximum, which is not only because there is more mass in the disc, but also that satellites are less likely to be lost into the planet and so more first generation satellites should survive.
	
	Fig. \ref{fig:IntMass1} shows that the integrated mass has corresponding behaviour: the increasing average mass means the distribution grows wider towards high masses. Meanwhile the peak moves towards lower mass and shrinks, as some of those satellites grow heavier in a bigger disc.
	
	\subsubsection{Radial distribution of satellites}
	\label{sec:pos2}
	
	In general, Fig. \ref{fig:Pos1} shows very similar behaviour for the 20 AU to 70 AU cases: a peak very close and a peak very far away and an evacuated middle section.
	The percentage of satellites close to the planet increases as disc size increases, because, as they on average migrate longer, it is easier for them to stop close to the planet instead of migrating into it. However, the percentage of satellites far out also increases, because with a bigger disc, the outer parts have more mass, so it becomes easier for an embryo to accrete a lot of mass.
	The middle section at 20 AU also shows a slight peak which vanishes as the disc size increases. As the disc grows in size, it will be harder for a satellite to move past the far peak and into this middle region. This is not only because the effective distance to migrate increases, but also because of what we established in Section \ref{sec:pos}, where we found that a satellite will need to accrete a lot of mass very early in its life in order to be able to migrate past the far peak. This becomes harder as the disc grows because the mass influx decreases and since satellites this far out tend to stay in a small region, because of the low gas density and migration, this means that their accretion time-scale is closely related to how fast the mass in their feeding zone is replenished.
	The 5 AU case is somewhat different than the other cases. The histogram of satellite locations does have a peak far out in the CPD, like the other cases, but it does not show a peak close to the planet. Instead, the peak in the middle region is more prominent. This is an effect of both the small disc size and the higher mass influx. The influx makes it easier for embryos to accrete mass fast and then migrate past the far peak. However, as the disc is small, embryos and satellites alike hardly stop when they are in the vicinity of the planet, rather will migrate into the planet more easily.
	
	\subsubsection{Formation temperature}
	\label{sec:formt2}

	The formation temperature in Fig. \ref{fig:Temp1} shows no big surprises. As you get closer to the host star, the disc heats up and thus the average temperature at which the surviving satellites are formed increases as well, albeit only slightly for the majority. It still shows that until 20 AU, most will form icy. The 5 AU case is a bit different, as this disc becomes hot enough in certain sections that dust evaporates. Thus there is a peak in formation temperature at around 1500 K, where dust evaporates. These will form late enough into disc lifetime that migrating into the planet will be very difficult. It also shows that in this case the satellites form significantly hotter in general.
	
	\subsubsection{Time-scale to reach the Europa-mass}
	\label{europa2}
	
	For all the cases, Fig \ref{fig:Eu1} shows that most of the satellites that reach the Europa mass scale do so between $10^4$ and $10^5$ years. However, the distribution widens towards fast time-scales as the disc grows bigger. There are two effects that play into this: first, since bigger discs have more mass, the average mass in the feeding zone of an embryo is higher, which allows for faster accretion. Second, as you get closer to the star, the smaller disc size means embryos that accrete mass very fast are more likely to migrate into the planet.
	We also see that the peak of the distribution moves towards longer time-scales as the disc grows bigger. That is because on average, the embryos will form farther out in the disc, so their orbital time-scale increases and thus their accretion time-scale increases as well. This means that, on average, the farther out an embryo is, the longer it would take it to reach Europa mass.
	 
    \subsection{Observational Predictions}
	
	\begin{figure}
			\centering
			\includegraphics[width=0.5\textwidth]{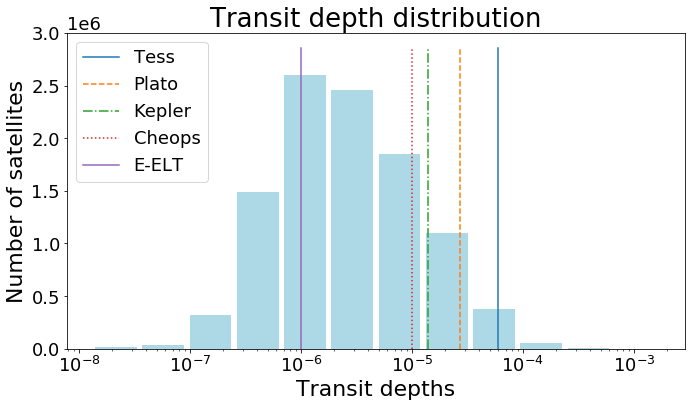}
			\caption{Plot of the transit depths for the satellites with at least Europa mass (any smaller ones will have a transit depth below all the equipment thresholds). The colored lines represent different instrument thresholds, therefore the population lying to the right of each line is that could be theoretically observed by the given instrument.}
			\label{fig:Depths}
	\end{figure}
	 
	Fig. \ref{fig:Depths} shows a plot of transit depths for every satellite created in the population synthesis with a mass greater than Europa. If they are smaller than that, they have a transit depth so low that no instrument could detect them, so we left those out of the plot. Even then, most of the satellites are too small to be detected for any instrument but E-ELT (before taking into account the geometrical factors). This already shows the difficulty of detecting exomoons such as those generated in the population synthesis. The averaged probability to find at least one satellite around an exoplanet confirms that, as shown in Table \ref{tab:DetectionProbTable}.
	
	\begin{table}
		\centering
		\caption{Detection probabilities for different instruments.}
		\label{tab:DetectionProbTable}
		\begin{tabular}{lccccr}
			\hline
			 & Kepler & Plato & Cheops & Tess & E-ELT\\
			\hline
			Probabilities & 1 \% & 0.5 \% & 1.3 \% & 0.15 \% & 3\% \\
			\hline
		\end{tabular}
	\end{table}
	
	Even with the best instrument, the probability of finding an exomoon around a given exoplanet is only around 3\%. This is not only because the satellites are of small mass, and hence size, but also because the heavier satellites tend to orbit close to the planet, which means that they spend more time in front of or behind it. For the satellites farther out it is the opposite; they tend to be smaller, but they also spend less time obscured by the planet.
	 
	\section{Discussion}
	\label{sec:discl}
	
	In the population synthesis approach several assumptions are needed to design a modeling framework simple enough to be flexible and allow studying the role and interplay of the various underlying physical processes. Such assumptions have a varying degree of impact on the results, which we discuss here.
	The first choice we made concerned the evolution of the disc. We started from gas density and temperature profiles that resulted from hydrodynamical simulations of \citet{Szulagyi16b} that generated a 10 $M_{\rm{Jupiter}}$ protoplanet and its CPD at 50 AU. We employed a simple model to evolve both density and temperature of the CPD exponentially on a certain time-scale, neglecting completely local thermodynamical effects.
	
	We also reduced the 3D disc structure extracted from the hydro simulation to construct a 1D model of the disc by using azimuthally averaged quantities, in accordance with \citet{Pringle81}. Parameters such as scale-height, surface density and sound speed,  extensively used in our model, are all born out of this simple 1D approximation. 
	
	Another choice we made concerns the generation of embryos. We used a seeding approach, where we set a certain set of randomly distributed locations at which we inserted the embryos. As a result, the creation of embryos was not tied to the disc parameters, apart from, implicitly, to the disc size.
	
	As we investigated the influence of the planet's semi-major axis, we rescaled the CPD properties (temperature, density) accordingly. For the temperature, we prescribed that the CPD has a higher or lower temperature normalization depending on the distance of the parent planet from the star. We also assumed that the form of the surface density profile does not change irrespective of the semi-major axis. We know that disc instability tends to generate gas giants from fragmentation predominantly in outer region of the CSD (at R > 30 AU), so if a planet is closer to the star, the more likely it e.g. migrated there from the CSD outer regions. The migration of the CPD, and the variation in temperature due to the varying stellar irradiation flux at varying distances would affect the density- and temperature profiles of the CPD significantly. Moreover, as the disc moves closer to the star, its Hill-radius would also shrink. Our results show that there are a lot of satellites in the outer regions, which could be stripped away as a result of this migration.
	
	There are also some parameters that we do not vary for simplicity, while we expect they should vary. One obvious example is the  initial seed mass, which we set at $10^{-8}$ planet masses. We tested varying that parameter, and
	determined that even lowering to as much as $10^{-10}$ planet masses has no significant influence on the results, because in the first few accretion cycles the embryos grow quickly to $~ 10^{-8}$ $ M_{\rm{planet}}$. Another possibility would be to determine the critical mass, that is the mass where the migration time-scale equals the accretion time-scale, at the position where an embryo is spawned. In the case of this disc, this mass is predominantly between ~$10^{-12}$ and ~$10^{-8}$ planet masses. In combination with our results where lowering the mass overall has no impact on the results, we chose the mass to be fixed for every embryo.
	
	We also set the seed locations at the beginning of the simulation and created later generations at the same locations instead of then varying it. We tested that using a different approach, namely randomly choosing the spawning locations at each subsequent iteration, does not affect the results.
	
	We also set the number of seeds between 5 and 20. We tested that increasing the maximum number does not have a big influence on the results, because it is extremely rare to have more than 10 massive satellites in the CPD at a given time. In principle, one could also change the maximum number of seeds depending on the disc size. As no disc every produces more than 10 massive satellites, 20 would be reasonable maximum in the 5 AU case, so it would increase from there for the cases where the disc is farther out. However, as we did test that increasing the number doesn't influence the results, we choose 20 as a maximum overall to save computation time.
	
	It is noteworthy that there is no other work on satellite formation in disc instability scenarios, nor works that address in general satellite formation around planets much more massive than Jupiter as in the present case,
	hence we could not compare directly with any previous work. To do that one would have to use the same model on two different discs, one CA and one GI, where the both of them were created in as much as possible the same environment. However, this is not the goal of this work, but we can
	still attempt a qualitative comparison with previous literature, which has exclusively focused on satellite formation in core-accretion and around a Jupiter sized planet, to address the formation of Galilean moons.
	\citet{Miguel16} used a minimum mass disc based on how the Jovian satellite system looks today with a cavity between disc and planet. A seeding approach was also used and accretion and migration of the embryos were taken into account, as well as dust and gas evolution. Temperature was not evolved and no dust influx from the stellar nebula was considered. The results showed a similar mass distribution as in our work, although with lower masses in general due to the disc being smaller and lighter. It also seemed to mirror our results for the positions of the orbits, with a cluster close to and a cluster far from the planet, although this may be influenced by the gas density which had a steep incline in the middle regions. The cavity also might influence how many satellites end up very close to the planet.
	
	\citet{Fujii17} used a more restrictive model to take a detailed look at the orbital evolution of moons. They evolved both surface density and temperature of their disc (around a 0.4 $M_{\rm{Jupiter}}$ planet) by numerically solving appropriate diffusion equations. They allowed their satellites to migrate and use a more extensive model of resonance trapping, and their satellites started with a fixed mass comparable to Io. They varied both influx and viscosity to test at different scenarios. Their results showed that, with the inclusion of proper resonance trapping, they could almost always get out a 1:2:4 resonance, that is present in the Galilean system. The resulting satellite positions ended up differently in their different runs, but the resonance was very consistently present. This is opposed to our model where we found no consistent resonant configurations. We tested this resonance model to our disc, which resulted in minor but noticeable changes. By and large, satellites will be slightly older, as some of the earlier heavier generations will survive, which is seen by an increase in maximum possible masses. It will not result in a significant increase in number of surviving satellites and only a slightly higher inner peak in the position plot and it also will not result in consistent resonant configurations. However, this model also was designed around a disc that has a region where migration based on gas-satellite interaction changes sign and thus effectively traps a satellite on a certain position, which will increase the chance of resonant configurations. As we do not have such a region in this disc, we decided against using this model in our work.
	
	\citet{cilibrasi18} used a model very similar to ours in terms of conceptual framework, but they developed that in context of a CPD formed by core-accretion, also drawn from a 3D radiative simulation as in our case. In their case, the gas giant was Jupiter-like both in mass as well as in orbital radius, which lead to a disc that smaller in size as well as mass and also hotter by an order of magnitude. Apart from the different
	CPD structure emerging in different formation scenarios, in the latter work a dust evolution model was used to determine the initial dust density, which naturally produced a region with increased density,  a so-called dust trap. The satellites were assumed to form there through streaming instability, which requires the local dust-to-gas ratio to be greater than unity. This alone leads to a very different mass distribution, very sharply peaked unlike the wide distribution in our work. That is because the satellite formation was tied to disc parameters and as a result it had to stop at a certain point. Since the satellites accreted most of their mass from the dust trap, very few remained close to the initial mass. Even though their disc was much lighter, their satellites still grew (relative to the planet mass) to comparable size as in our case, even a bit more massive. This can also be attributed to the dust trap, as  the dust mass was heavily concentrated there, and on top of that, the influx would be high in that region as well, as the satellites would predominantly accrete mass in this region. So even though there are stark differences in disc parameters, the mass (at least in terms of planet masses) of satellites seem to be comparable, if with a different distribution. Another notable similarity are the formation temperatures; even though their disc (being much closer to the sun it orbits) was an order of magnitude hotter, still produced mostly icy satellites. This means that the surviving satellites were very young, which is also reflected in the fact that in their case more mass is lost into the planet, which translates to more satellites migrated into the planet. The comparison in number of surviving satellites is also interesting: even though the disc is smaller and satellites are on average being create closer to the planet in their case, most systems still have 3 or 4 massive satellites, while in our case there are mostly 0, 1, or 2. This can again be attributed to the dust trap and the younger age, with lets them accrete much mass while also not being subject to fast migration.
	
    Even though the similarity between the two models allow us yo roughly compare them, it is hard to quantify how many of these differences and similarities are a result of the different planet formation scenarios (CA or GI), their mass and their orbital position. \citet{Szulagyi16b} attempted to do a direct comparison of gas giants created by both CA and GI. Their results show that a CA gas giant created at 50 AU with a mass of 10 $M_{\rm{Jupiter}}$ would have a CPD that is about an order of magnitude hotter, about 12 \% as massive as our CPD and less dense. This should in principle lead to satellites that are less massive and created in a hotter environment, but comparing this is not within the scope of this work.
	
	\section{Conclusions}
	\label{sec:concll}
	
	In this work we investigated the formation of satellite systems around a 10 $M_{\rm{Jupiter}}$ gas giant created by gravitational instability at a semi-major axis of 50 AU. We used the gas density and temperature profiles as well as the mass influx of a circumplanetary disc obtained with an SPH simulation \citep{Szulagyi16b} and assumed that the dust had the same profile as the gas, but multiplied by the overall dust-to-gas ratio of the disc. We then used a population synthesis approach, where we placed satellite embryos in the circumplanetary disc. The disc itself was also evolved over time on the dispersion time-scale and refilled with micron-sized dust as well from the meridional circulation. Within the population synthesis framework, the embryos migrated, accreted mass and collided while also allowing for subsequent satellite generations to be formed.
	
	As the population synthesis approach requires, we randomized the not well constrained/not constant initial conditions: the dust-to-gas ratio, the disc dispersion- and refilling time-scales as well as the number and initial positions of the seeds at the beginning of every run. We also investigated the influence of the semi-major axis of the planet (and hence its CPD) on the resulting satellite system architecture by re-scaling the CPD's size, temperature and mass influx with the semi-major axis.
	
	In the nominal case (planet with a CPD at 50 AU from the star), our results showed that in $\sim$ 60 \% cases,  1--4 satellites formed with a mass of minimum $10^{-5}$ planet masses, similarly to what we can observe in our Solar System's satellite systems. Most of the satellites have masses in the range of the Jovian satellites, but there is a significant amount that have a higher mass than that, with the maximum being around $10^{-3}$ planet masses (i.e. 3 Earth masses). The integrated mass had a peak at around $2 \cdot 10^{-5}$ planet masses, an order of magnitude lower than what we can observe in the case of Jovian, Saturnian and Uranian moon systems. A lot of satellites ended up migrating into the planet, leading to an average of $\sim$10 Earth-masses of solids "polluting" the planet's atmosphere and envelope, increasing its overall metallicity. With sedimentation, some of this material could potentially make a small solid core inside this GI formed gaseous planet.
	
	Contrary to previous satellite formation studies, we also found that there is a significant amount of massive satellites with an orbital radius around ~0.5 to 0.6 $R_{\rm{disc}}$. This is a result of the massive CPD in our model, which allows for seeds in that region to grow that massive. Because of the low dust densities in that region, the increase is slow enough to not overcome the decrease in gas density, which leads to a slow migration. The vast majority of satellites were created in the CPD at temperatures much lower than the freezing point of water and as such these moons are very likely to be icy in composition. This is no surprise, as the temperature of the disc was very low to begin with. Most satellites also formed  between $10^4$ and $10^5$ years, comparable to the gas disc dispersion time-scale, with some having formed even very rapidly in around $10^3$ years. 
	
	The influence of the semi-major axis of the planet on the number of satellites and their formation time-scales turned out to be small. The mass histogram of moons showed that it becomes harder for heavy satellites to form in CPDs closer to the star. First, because they could more easily migrate into the planet and second, because the overall CPD mass is smaller to make moons.
	
	We also found that the probability of detecting satellites like the ones in this population synthesis is very low ($\leq$ 3\%) even with E-ELT. This is mainly because the satellites are expected to form beyond the snowline (around giant and ice giant planets), and the current detection methods are not sensitive to this further-out population. On top of this, the heavier satellites tend to orbit close to the planet which makes their detection more challenging. However, giant planets can get closer to the star via various dynamical evolution processes, and if they could keep their moons during this process, then those satellites should be easier to detect. 
	
	\section*{Acknowledgements}
    This work has been carried out within the Swiss National Science Foundation (SNSF) Ambizione grant PZ00P2\_174115.  We thank for the anonymous referee for their suggestions to improve our paper.
    
   \section*{Data availability}
    The data underlying this article will be shared on reasonable request to the corresponding author.
	
	
	
	


	


	\bsp	
	\label{lastpage}
\end{document}